\documentclass{ptephy_v1}

\preprintnumber{XXXX-XXXX} 

\usepackage{url}
\usepackage{braket}
\usepackage{textcomp}
\usepackage{mathcomp}

\begin{document}

\title{Response of the underground environment of the KAGRA observatory against the air pressure disturbance from the Tonga volcano eruption on January 15th, 2022}


\author[1,*]{Tatsuki Washimi}
\author[2]{Takaaki Yokozawa} 
\author[3]{Akiteru Takamori} 
\author[3]{Akito Araya}
\author[4]{Sota Hoshino}
\author[5]{Yousuke Itoh} 
\author[5]{Yuichiro Kobayashi} 
\author[6,7]{Jun'ya Kume} 
\author[2]{Kouseki Miyo} 
\author[4]{Masashi Ohkawa} 
\author[2]{Shoichi Oshino} 
\author[1]{Takayuki Tomaru} 
\author[6,7]{Jun'ichi Yokoyama} 
\author[2]{Hirotaka Yuzurihara} 

\affil[1]{Kamioka Branch, National Astronomical Observatory of Japan (NAOJ), Kamioka-cho, Hida City, Gifu 506-1205, Japan}
\affil[2]{Institute for Cosmic Ray Research (ICRR), KAGRA Observatory, The University of Tokyo, Kamioka-cho, Hida City, Gifu 506-1205, Japan}
\affil[3]{Earthquake Research Institute, The University of Tokyo, Bunkyo-ku, Tokyo 113-0032, Japan}
\affil[4]{Graduate School of Science and Technology, Niigata University, Nishi-ku, Niigata City, Niigata 950-2181, Japan}
\affil[5]{Department of Physics, Graduate School of Science, Osaka Metropolitan University\footnote{The name was changed from \textit{Osaka City University} on April 1st, 2022}, Sumiyoshi-ku, Osaka City, Osaka 558-8585, Japan}
\affil[6]{Research Center for the Early Universe (RESCEU), The University of Tokyo, Bunkyo-ku, Tokyo 113-0033, Japan}
\affil[7]{Department of Physics, The University of Tokyo, Bunkyo-ku, Tokyo 113-0033, Japan}
\affil[*]{Corresponding author}


\begin{abstract}
On January 15, 2022, at 04:14:45 (UTC), the Hunga Tonga-Funga Ha'apai, a submarine volcano in the Tongan archipelago in the southern Pacific Ocean, erupted and generated global seismic, shock, and electromagnetic waves, which also reached Japan, situated more than 8,000 km away. 
KAGRA is a gravitational wave telescope located in an underground facility in Kamioka, Japan. It has a wide variety of auxiliary sensors to monitor environmental disturbances which obstruct observation of gravitational waves. The effects of the volcanic eruption were observed by these environmental sensors both inside and outside of the underground facility. In particular, the shock waves made it possible to evaluate the transfer functions from the air pressure wave in the atmosphere to the underground environmental disturbances (air pressure and seismic motion).
\end{abstract}

\subjectindex{H, F30, H20, J63}

\maketitle

\section{The KAGRA gravitational wave observatory and its environmental monitoring systems}\label{sec:KAGRA}
\begin{figure}[!h]\centering
    \includegraphics[width=15cm]{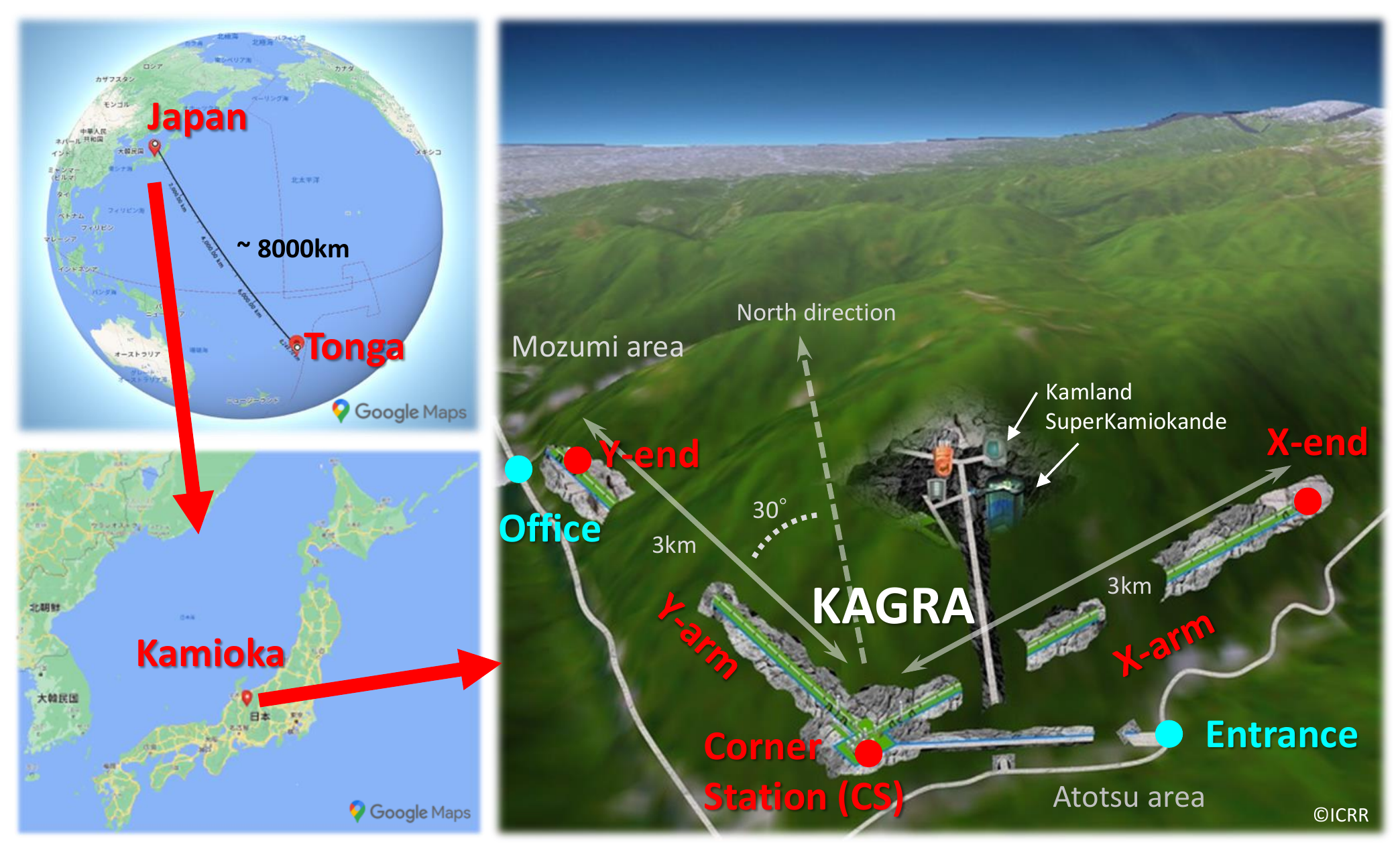}
    \caption{An image of the KAGRA experimental site and the names of each location.}
    \label{fig:KAGRA}
\end{figure}

KAGRA~\cite{PTEP01, Galaxies} is a laser interferometric gravitational wave detector with two arms, 3~km long, constructed in Kamioka, Japan. In contrast to other gravitational wave detectors, such as LIGO~\cite{LIGO}, Virgo~\cite{Virgo}, and GEO~\cite{GEO}, KAGRA has been built in an underground facility to avoid and filter noise sources caused by human activities, seismic disturbances, and meteorological phenomena. For example, seismic noise can be reduced by approximately a factor of $\mathcal{O}(10^2)$ at underground facilities compared with the surface above 1~Hz~\cite{Tomaru}. 
Figure~\ref{fig:KAGRA} shows a schematic of the KAGRA experimental site. The X- and Y-arms of KAGRA are rotated $-30\tcdegree$ in the east and north directions, respectively. The Z-direction is defined as the vertically upward direction. The underground depth in vertical is 200~m for the corner station (CS) and 450~m for the X- and Y-end. One access tunnel connects the entrance in the Atotsu area to the CS and the other connects the outside in the Mozumi area to the Y-end station. There is no access tunnel nor ventilation shaft to the X-end and the fresh air is provided through the X-arm tunnel. 
The office building is located in the Mozumi area, outside the tunnel. 

As a gravitational wave detector is highly delicate and the targeted signals are very small, it is extremely important to monitor and study the environmental disturbances quantitatively and exhaustively, despite the underground facility being quieter and more stable relative to the surface.
Physical environmental monitoring (PEM) for the gravitational wave observation involves monitoring such noise, utilizing seismometers, accelerometers, microphones, magnetometers, \textit{etc.}. Details of the KAGRA PEM are reported in the cited literature~\cite{PTEP3, Galaxies}. 
The target frequency of gravitational wave observation is 5--5000~Hz~\cite{PTEP01}, and it is necessary to mitigate any noise in this range. Although frequencies lower than 1~Hz are not used in the current astrophysical analysis, this frequency band is also important for performing the interferometer operation. This is because the distance between the mirrors of the interferometer must be maintained to an accuracy of a few hundred pico-meters for the resonant state. Low-frequency vibrations ($<20$~Hz) affect the control of the suspensions and interferometer. 

\begin{figure}[!h]\centering
    \includegraphics[width=12cm]{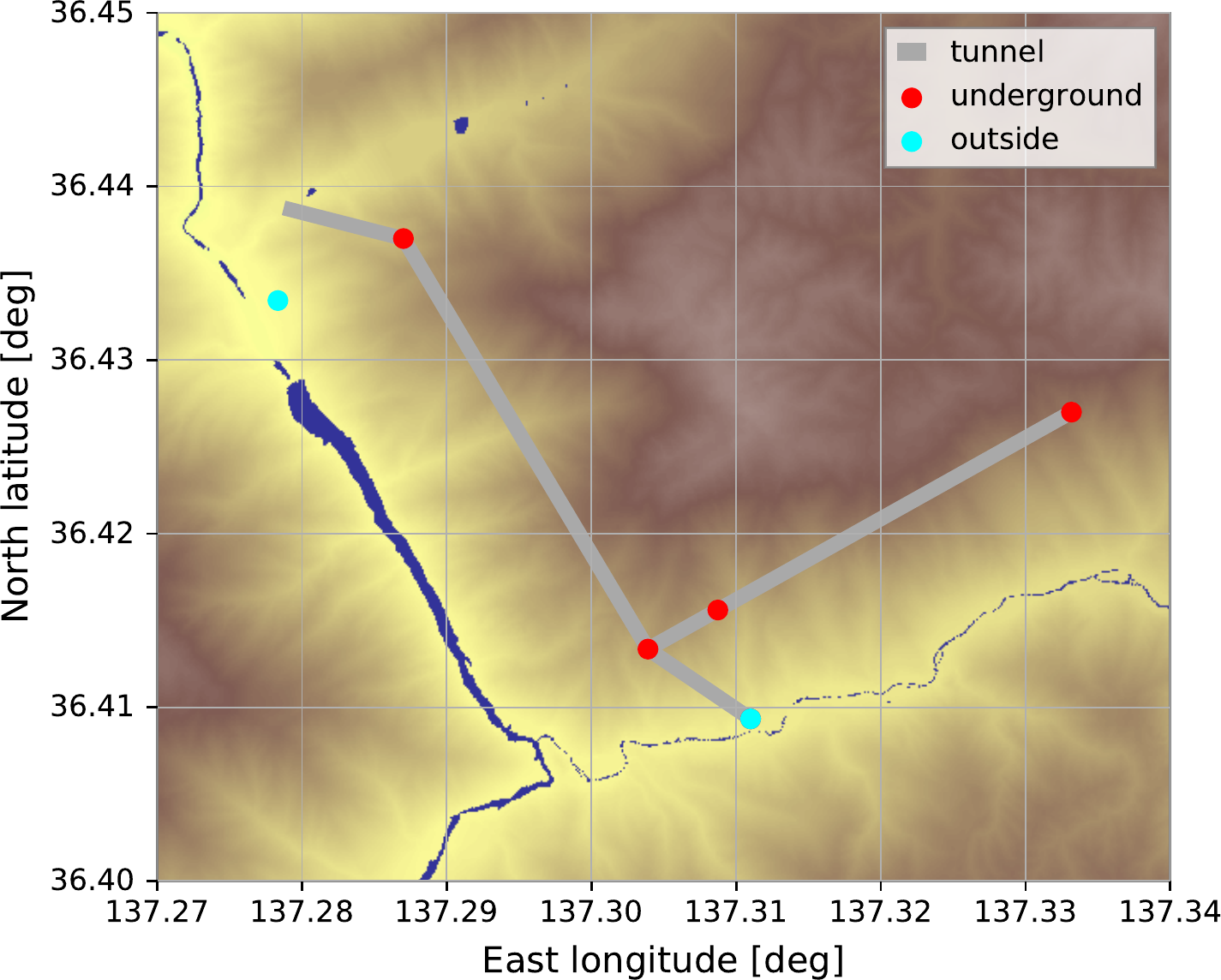}
    \caption{Longitude and latitude for each location. The red markers are underground (CS, X-arm 500~m, X-end, and Y-end) and the cyan markers are outside (office at Mozumi, entrance at Atotsu).}
    \label{fig:LonLat}
\end{figure}

The environmental sensors used in this study are summarized in Table~\ref{tab:sensors}, and the locations of the experimental areas are shown in Fig.~\ref{fig:KAGRA} and Fig.~\ref{fig:LonLat}, with red markers for the underground and cyan markers for the outside of the tunnel.

\begin{table}[!h]\centering
\caption{List of the environmental sensors used in this study.}
\label{tab:sensors}
\begin{tabular}{cccc}\hline
Sensor type & Product name                & Catalog $f$ range     & Location \\ \hline
Seismometer & Nanometrics Trillium 120QA  & 10~mHz--10~Hz        & CS, X-end, Y-end\\ 
Microphone  & ACO 4152N                   & 20~Hz--20~kHz        & CS \\
Infrasound sensor & SAYA INF01LE          & 0.3~mHz--6~Hz        & Office \\
Barometer   & SAYA INF01LE                & DC--1~Hz             & Office \\
Barometer   & Davis Vantage Pro2          & DC (1~min/S)          & Entrance \\
Barometer   & Vaisala BAROCAP PTB110      & DC                    & X-arm 500~m\\
Magnetometer& Bartington Mag-13MCL100     & DC--3~kHz            & Y-end \\ \hline
\end{tabular}
\end{table}

\section{Environmental signals from the Tonga volcano eruption observed at the KAGRA observatory}\label{sec:Tonga}
On January 15, 2022, at 04:14:45 (UTC), the undersea volcano of Hunga Tonga-Funga Ha'apai erupted (lat. ~20.546\textdegree S, lon.~175.390\textdegree W, $\pm 8.1$~km, and VEI\footnote{Volcanic Explosivity Index}~6)~\cite{USGS, GVP, Klein, Zhao}. 
The impulsive shock wave caused by the volcanic eruption transformed into long-period air pressure waves and were observed worldwide. 
These waves also reached Japan, travelling more than 8,000~km from the Tonga volcano, over the Pacific Ocean, and were observed as disturbances in the air pressure ~\cite{KUT, Kataoka, Saito}. 

\begin{figure}[!h]\centering
    \includegraphics[width=15cm]{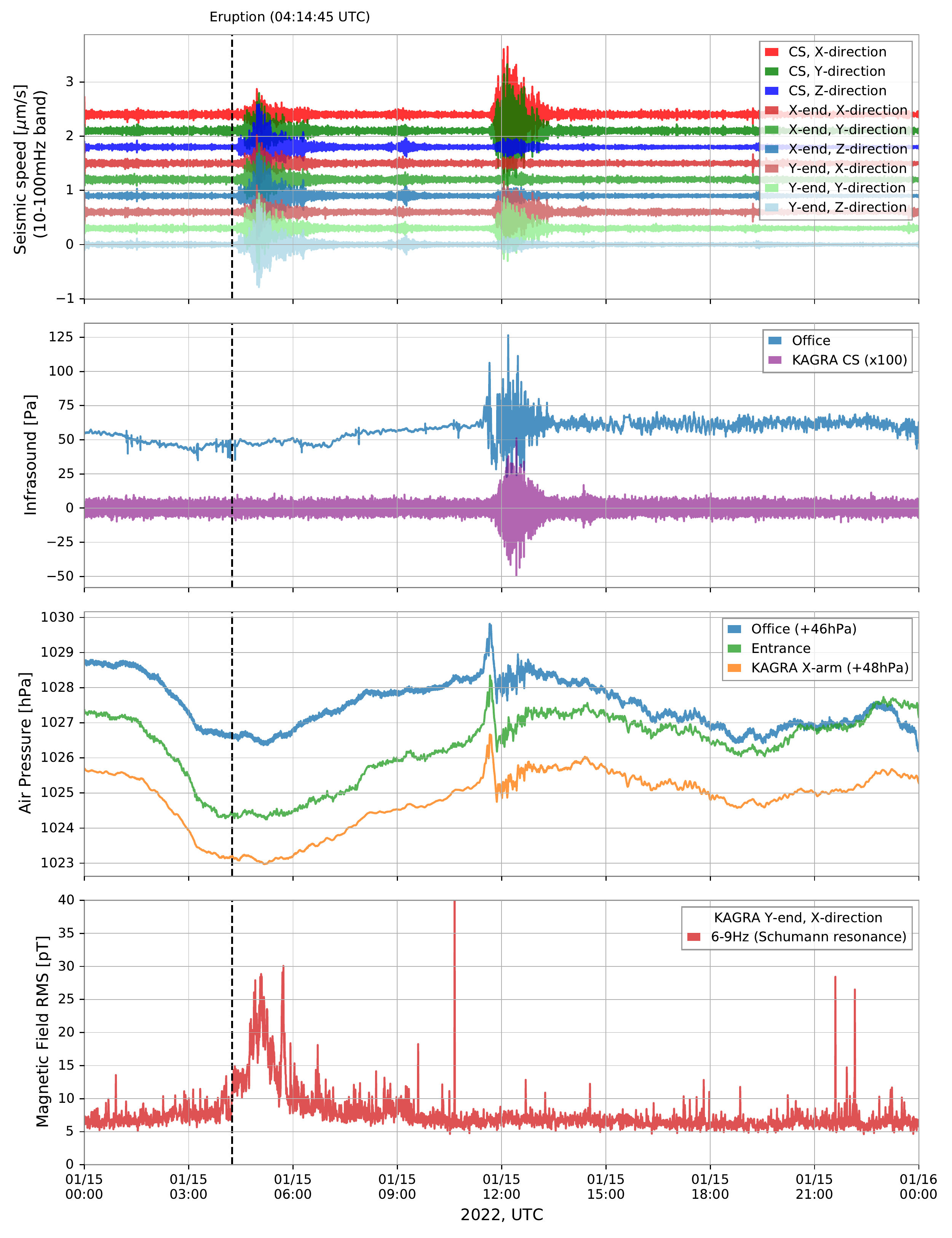}
    \caption{The summary of the eruption signals monitored by the environmental sensors of KAGRA. 
    1st panel: The seismic speed with 10-100~mHz band-pass filter monitored by the seismometers inside the tunnel. 
    2nd panel: Infrasound monitored outside and inside of the tunnel. 
    3rd panel: Absolute air pressure monitored outside and inside of the tunnel. 
    4th panel: 6--9~Hz band limited RMS for the magnetometer at the Y-end.}
    \label{fig:Summary}
\end{figure}

These signals emanating from the volcanic eruption were observed by the environmental sensors of KAGRA, not only outside the tunnel, but also within the underground facility. 
Figure~\ref{fig:Summary} summarizes the eruption signals monitored by the environmental sensors of the KAGRA. Clear signals were detected by the magnetometer moments after the eruption, by the seismometers several minutes later, and subsequently in the seismometers, infrasound sensors, and barometers, approximately 7~h later. The details of each sensor and its detected signals are presented in this section, and a combined analysis of the sensors is discussed in the next section.

\subsection{Seismic signals}
\begin{figure}[!h]\centering
    \includegraphics[width=14cm]{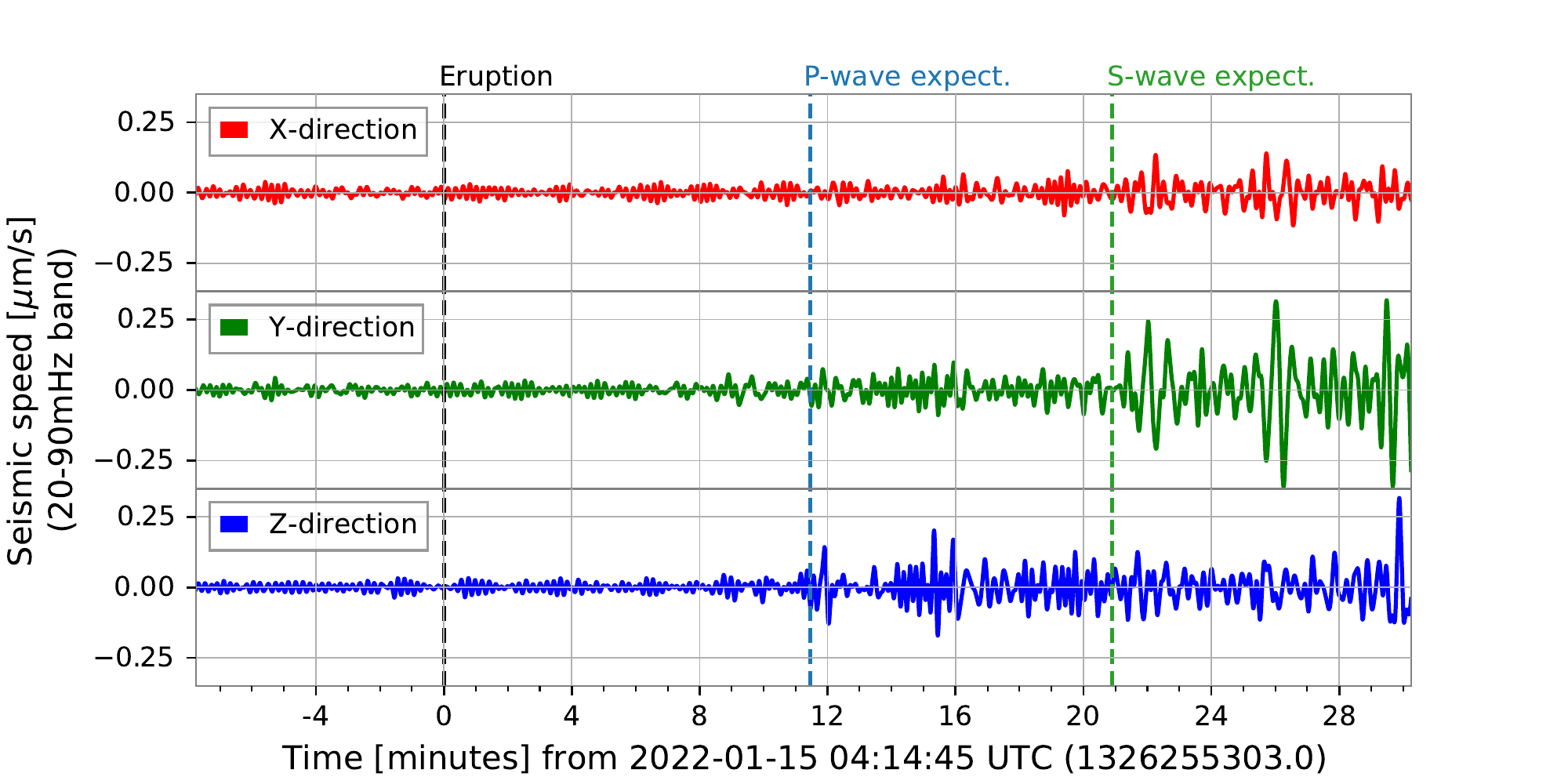}
    \caption{Zooming up for the seismic wave signals, average of the three locations for each direction).}
    \label{fig:EQzoom_CS}
\end{figure}

Tri-axial seismometers (Nanometrics Trillium 120QA~\cite{T120QA}) monitoring seismic speeds are located at the CS, X-end, and Y-end inside the tunnel. Their axes are aligned to the coordinates of KAGRA, resulting in a total of nine seismic channels.
Figure~\ref{fig:EQzoom_CS} shows a magnified view of the time series of the seismometers (average of the three locations for each direction), with the 20--90~mHz band passed. 
Regarding the seismic wave signal, motion along the Z-direction was detected approximately 12~min after the eruption and that along the X/Y-direction approximately 22~min after the eruption. 
The arrival times of the $P$-wave and the $S$-wave were expected to be 11~min and 21~min, respectively, as calculated by \texttt{Obspy~1.3.0} with the \texttt{iasp91} model~\cite{ObsPy}. These values are consistent with the observed values. The observed $S$-wave is smaller in vertical compared with horizontal as expected, because it is a transverse wave and comes from downside to upside. The typical frequency of the seismic body wave from this eruption is reported to be 30--80~mHz~\cite{U09-P29}, which is much lower than that of typical earthquakes (order 1~Hz). 
Thus, it is confirmed that the seismic signals observed at the KAGRA site originated from the eruption. 

\begin{figure}[!h]\centering
    \includegraphics[width=13cm]{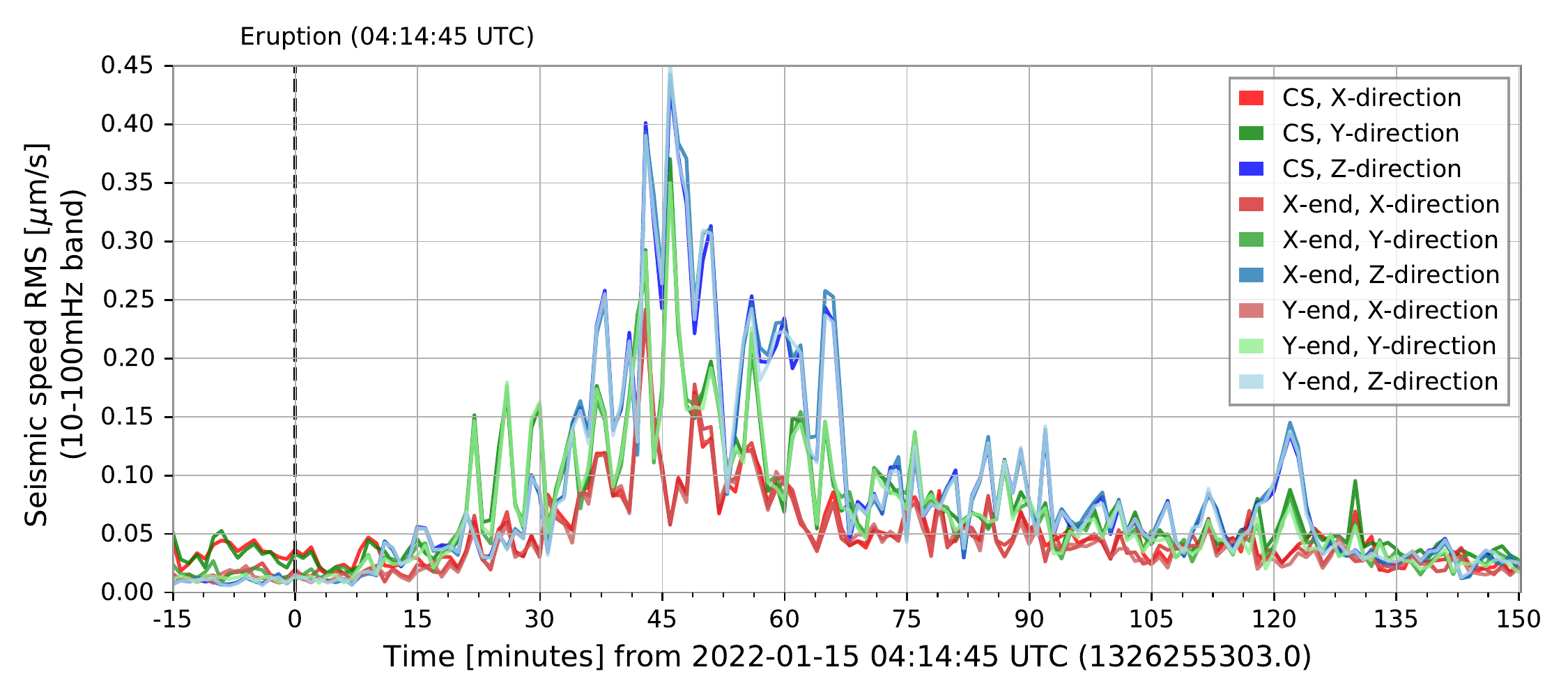}
    \includegraphics[width=13cm]{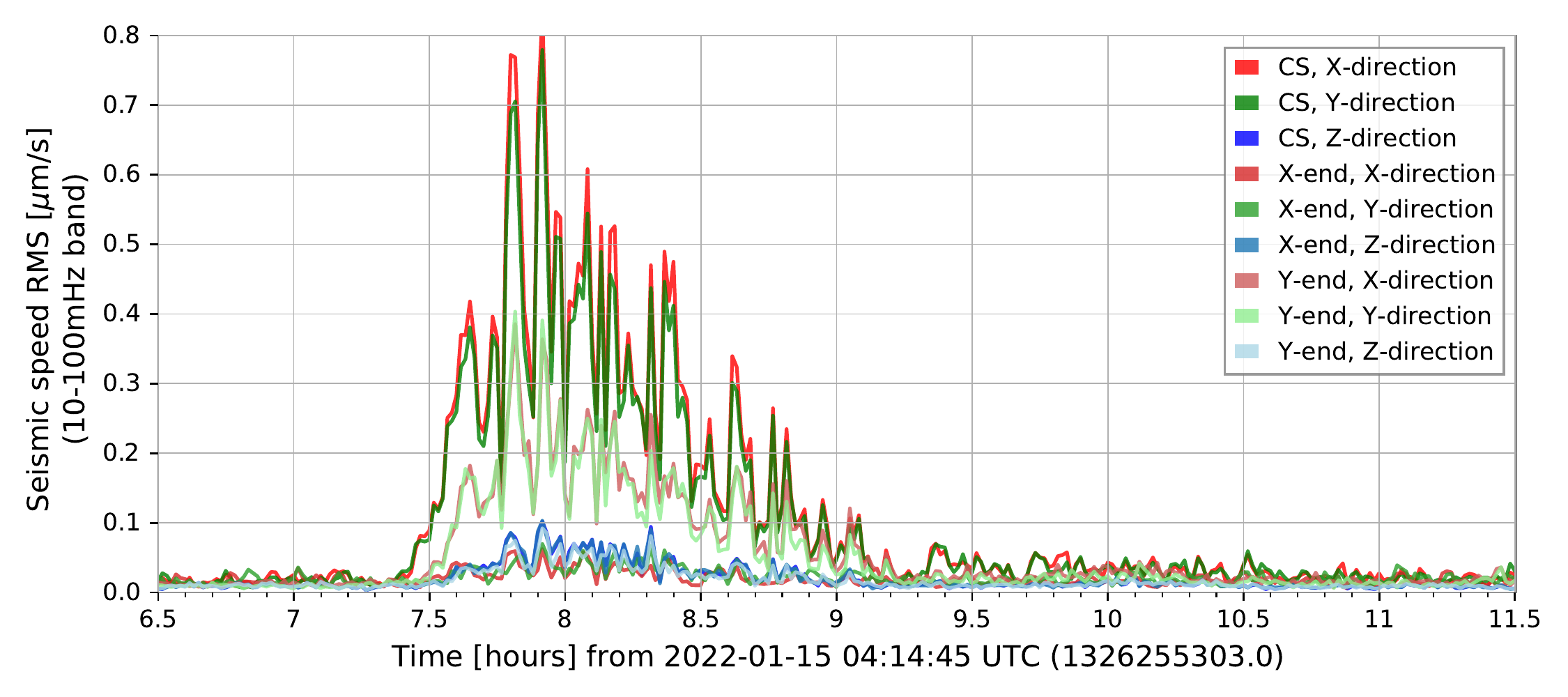}
    \caption{The time series of the 10-100~mHz band limited RMS for every 1 min for each channel of the seismometer. Top: the seismic wave, Bottom: the atmospheric wave.}
    \label{fig:SEIS_RMS}
\end{figure}

The time series of the 10-100~mHz band-limited RMS for every 1min for each channel of the seismometers is shown in Fig.~\ref{fig:SEIS_RMS}. The top and bottom parts focus on the time windows for seismic waves and air pressure waves, respectively. $t=0$ on the horizontal axis represents the eruption time. The red, green, and blue colored lines represent the X-, Y-, and Z-directions, respectively. Regarding the brightness, the deep, intermediate, and light colors represent the CS,  X-end, and Y-end locations, respectively. The seismic wave signals were almost the same in each direction, independent of the location. The air pressure wave signals exhibited properties different from those of the seismic waves. They significantly varied among the locations and were similar in the horizontal directions, X- and Y-, at each location.

\begin{figure}[!h]\centering
    \includegraphics[width=15cm]{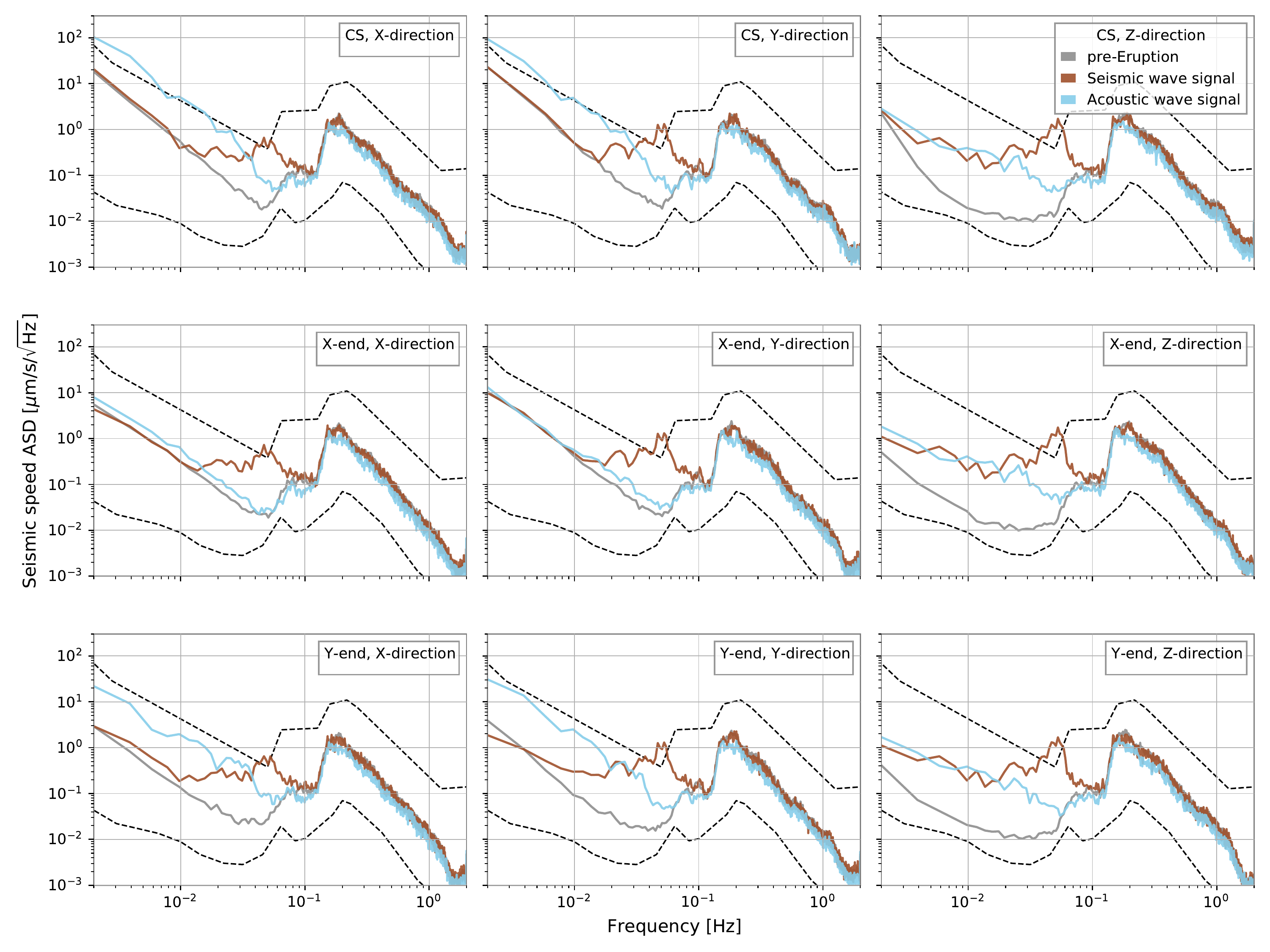}
    \caption{The ASDs of the seismometers for the background data (gray), the seismic wave signal (brown), and the air pressure wave signal (sky blue). Dotted lines are Peterson's NHNM and NLNM~\cite{Peterson}.}
    \label{fig:SEIS_ASD}
\end{figure}

Figure~\ref{fig:SEIS_ASD} shows the amplitude spectral densities (ASDs) of the seismometers for three time windows: pre-eruption (as background data), seismic wave signal, and air pressure wave signal. 
The frequency response of this seismometer (see Appendix~\ref{sec:Trillium120}) has been corrected.
Their UTC times are summarized in Table~\ref{tab:ASDtime}. 
The conspicuous peak at 0.2~Hz is the micro-seismic motion induced by the ocean waves, which is independent of the eruption signal. The seismic signal (soil brown lines) has a peak at approximately 0.05~Hz, and the air pressure signal (sky blue lines) has a distribution below 0.07~Hz. 

\begin{table}[!h]\centering
\caption{UTC time for calculating the ASDs in Fig.~\ref{fig:SEIS_ASD}.}
\label{tab:ASDtime}
\begin{tabular}{lcc}\hline
Timing               & Start               & End  \\ \hline
pre-Eruption         & 2022-01-15 00:00:00 & 2022-01-15 04:14:45 \\ 
Seismic wave signal  & 2022-01-15 04:35:00 & 2022-01-15 06:00:00 \\ 
Acoustic wave signal & 2022-01-15 11:30:00 & 2022-01-15 13:30:00 \\ \hline
\end{tabular}
\end{table}

\subsection{Air pressure signals}
Infrasound includes acoustic waves in the lower frequency range less than 20~Hz, and below the human hearing range (20~Hz to 20~kHz). It can be used to understand various geophysical phenomena, such as volcanic eruptions, tsunamis, earthquakes, landslides, meteorites entering the atmosphere, and various man-made noises (wind power generation, explosions, nuclear tests, \textit{etc.}). 
The infrasound sensor (SAYA INF01LE~\cite{SAYA, SAYA-INF01LE}) operates in an office building on the ground surface. It can monitor the infrasound in the frequency range of 0.3~mHz--6~Hz, and dynamic range of 134.2~dB ($\pm 733.4$~Pa) with a resolution of 0.19~mPa and low-noise (7~mPa rms). It also has a barometer for monitoring the absolute pressure.

\begin{figure}[!h]\centering
    \includegraphics[width=15cm]{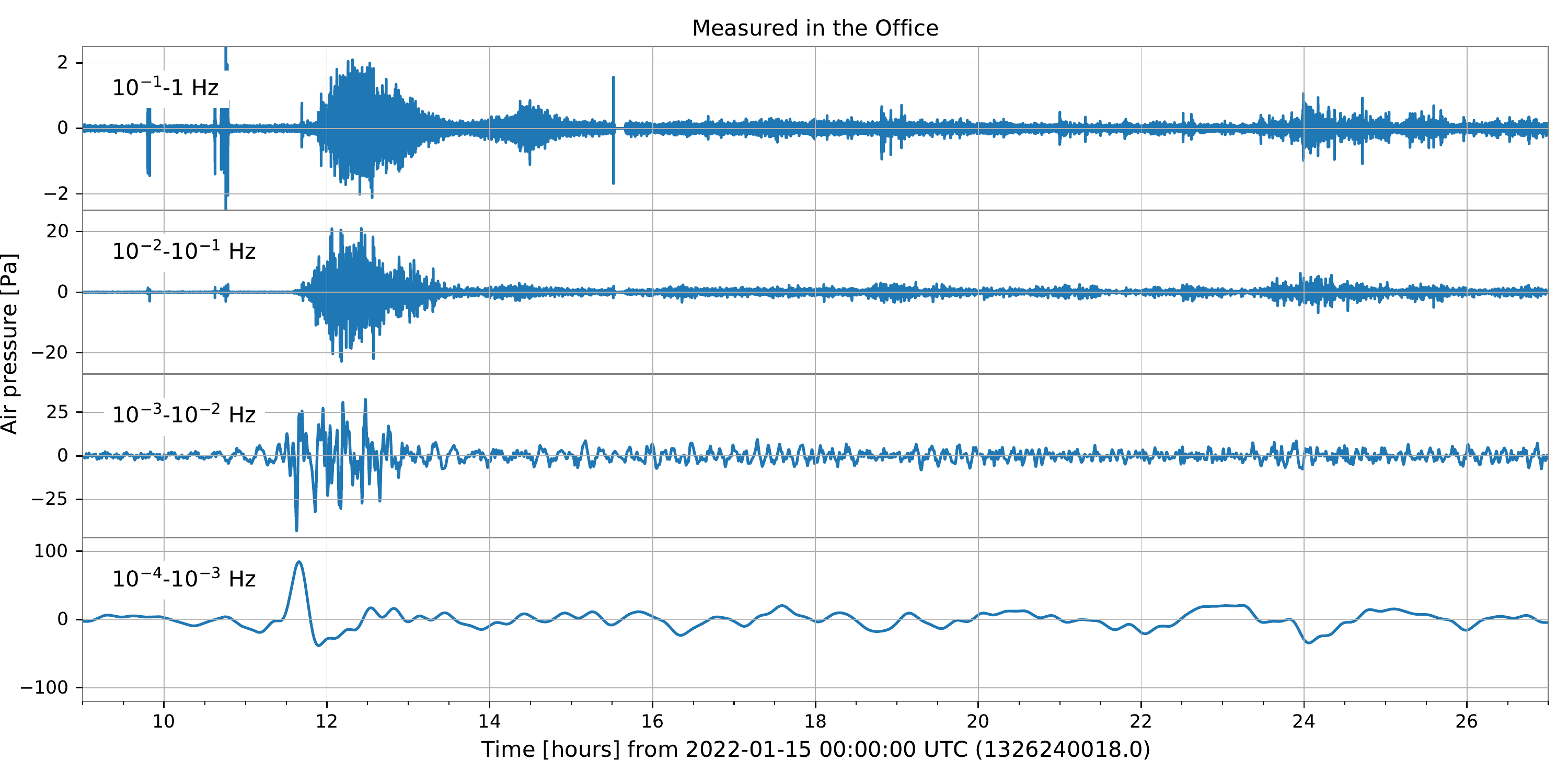}
    \caption{The band-limited time series of the infrasound sensor (1st and 2nd panels) and the barometer (3rd and 4th panels) of the SAYA INF01LE located in the office.}
    \label{fig:INF01_TimeSeries}
\end{figure}

\begin{figure}[!h]\centering
    \includegraphics[width=11cm]{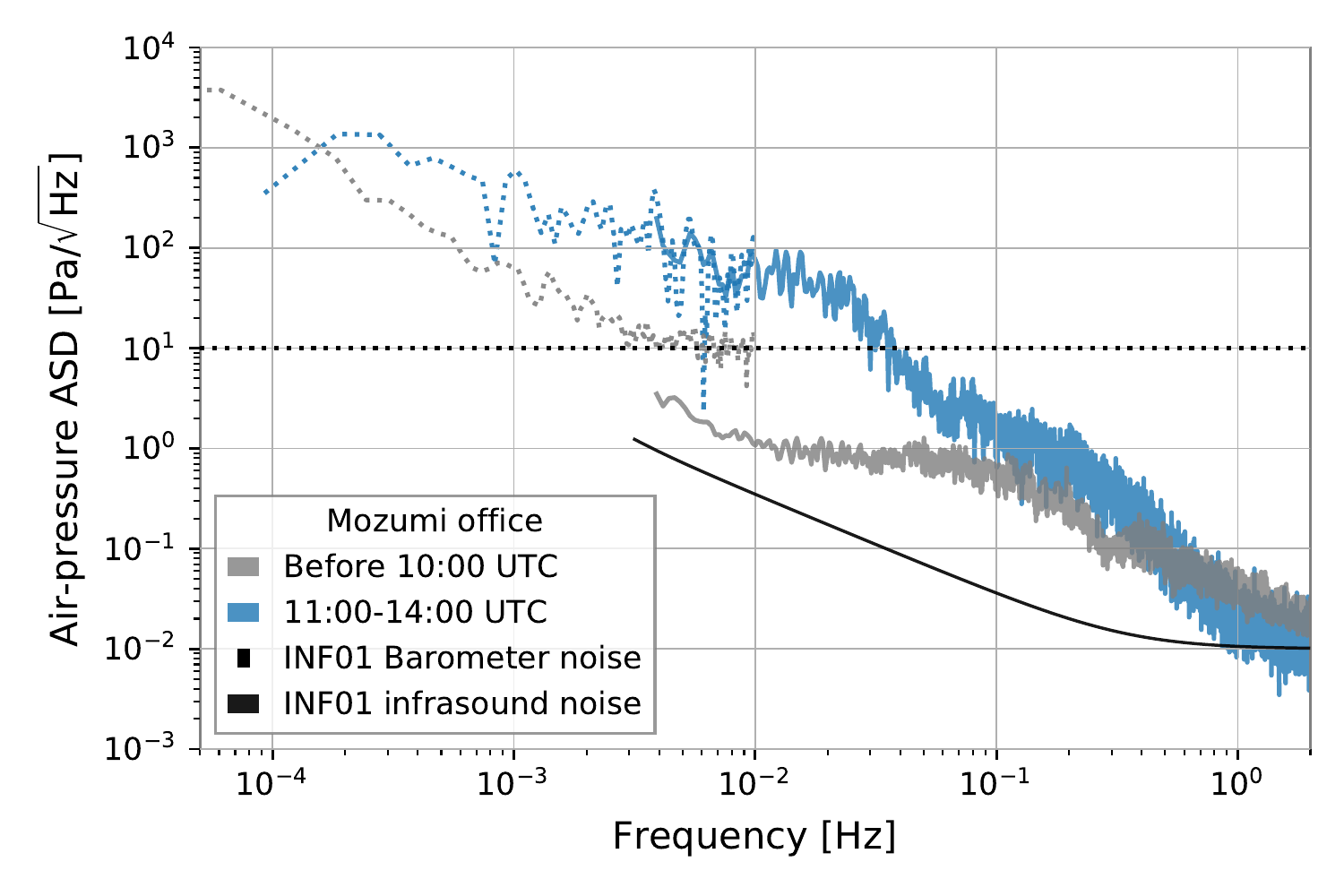}
    \caption{The ASDs of the SAYA INF01LE located in the office. solid lines : infrasound sensor, dotted lines : barometer. Gray : background, blue : the eruption main signal, black : sensor noise.}
    \label{fig:INF01_ASD}
\end{figure}

The band-limited time series of the infrasound sensor ($10^{-1}$--1~Hz band and $10^{-2}$--$10^{-1}$~Hz band) and barometer ($10^{-3}$--$10^{-2}$~Hz band and $10^{-4}$--$10^{-3}$~Hz band) of the INF01LE located in the office are shown in Fig.~\ref{fig:INF01_TimeSeries}. Their spectrogram is shown in Fig.~\ref{fig:INF01_ASD}.
The Lamb waves of the Earth's atmosphere were observed around 11:30 UTC, and an atmospheric gravity wave in the milli-Hz range was observed for a long time. This is consistent with other reports from Japan~\cite{KUT, Kataoka, Saito}.
The small transient disturbances in the higher frequency bands that appeared before the arrival of the air pressure wave signal are understood to be caused by human activities, such as opening/closing doors in the building.

\begin{figure}[!h]\centering
    \includegraphics[width=12cm]{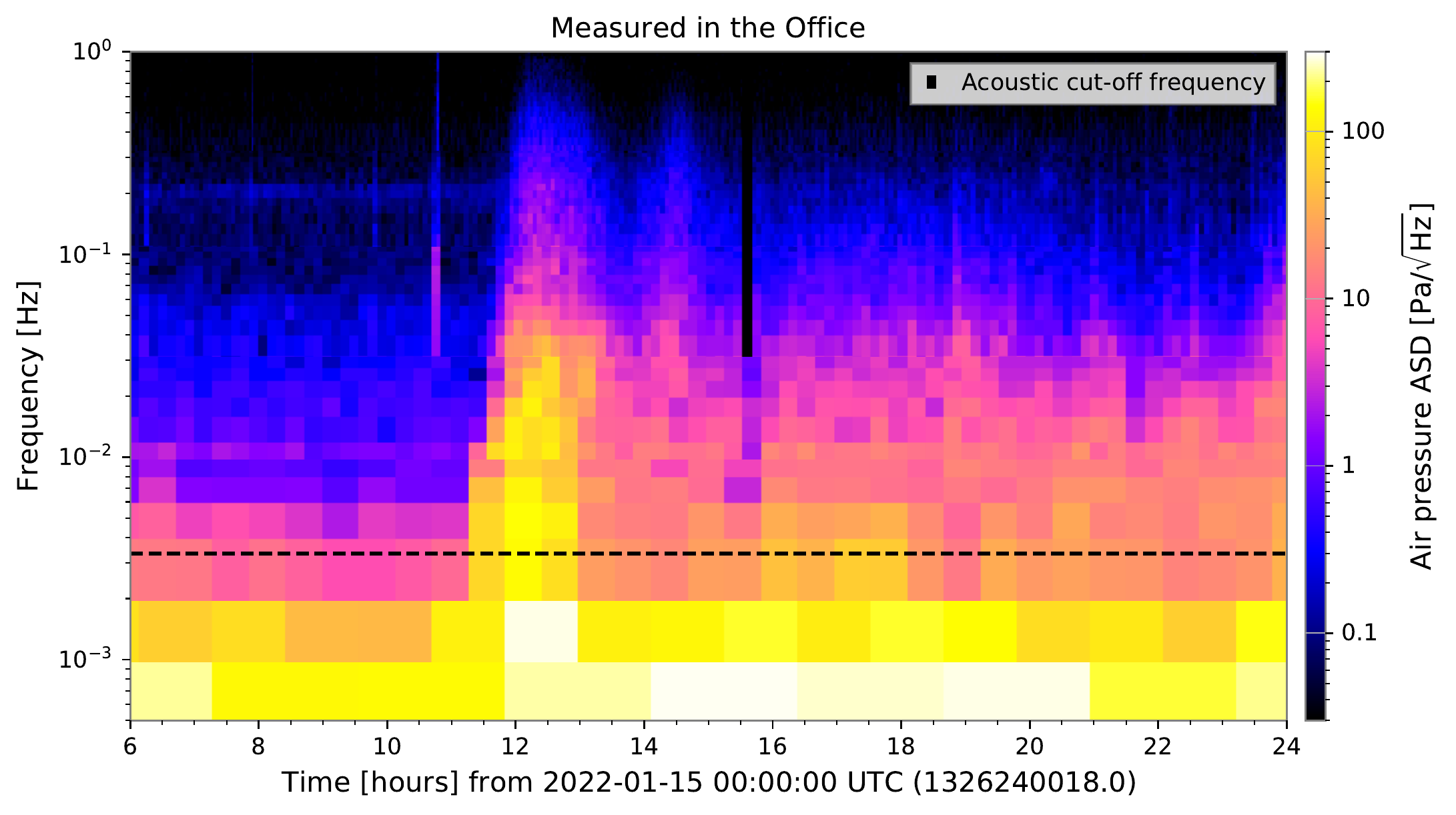}
    \caption{The spectrogram of the SAYA INF01LE located in the office.}
    \label{fig:INF01_SG}
\end{figure}

Figure~\ref{fig:INF01_ASD} shows the ASDs of the infrasound sensor (solid lines, over $3\times10^{-3}$~Hz) and barometer (dotted lines, below $1\times10^{-2}$~Hz) of the INF01LE in the office. Each color corresponds to the time of the background (gray) and the eruption’s main signal (blue). 
For the infrasound sensor, the eruption signal was detected below 0.5~Hz larger than the background noise. 
The ASD of the barometer was limited by the sensor noise at 10~Pa/$\sqrt{\mathrm{Hz}}$ for a background signal above $3\times10^{-3}$~Hz. However, the eruption signal was larger than the background signal over $10^{-4}$~Hz, which was consistent with the infrasound sensor at approximately $10^{-2}$~Hz. 
In the following analysis, the infrasound sensor signal used was over $1\times10^{-2}$~Hz, and the barometer signal, below this frequency, as the air pressure signal measured in the office.

Two other barometers were operating at the entrance of the KAGRA tunnel (1~min sampling) and at the X-arm (1~sec sampling), 500~m from the CS. 
The time series of these barometers are plotted in the 3rd panel of Fig.~\ref{fig:Summary}. A transient pressure disturbance of 2~Pa was observed in all barometers, which is consistent with numerous other reports in Japan~\cite{KUT, Kataoka, Saito}.

\medskip

At the underground experimental site of KAGRA, there are about 20~microphones, which are sensitive to audible sound (20~Hz--20~kHz). One of them, located in the CS, is connected to a low-frequency amplifier and has a sensitivity range down to approximately 10~mHz. 
During the gravitational wave observation run of KAGRA, the correlation between the low-frequency microphone and the gravitational wave channel caused by the infrasound from the wind excitation and disturbances outside the tunnel has been reported~\cite{CAGmon}.
The frequency response of this system (microphone and amplifier) applied to the following analysis was provided by the manufacturing company, as summarized in the appendix.~\ref{sec:TF_ACO4152NHA}.

Figure~\ref{fig:infrasound} shows the time series of the infrasound sensor at the office (blue) and in the KAGRA CS (purple) with a 0.05--0.5~Hz bandpass filter. The difference was approximately a factor of 10.

\begin{figure}[!h]\centering
    \includegraphics[width=12cm]{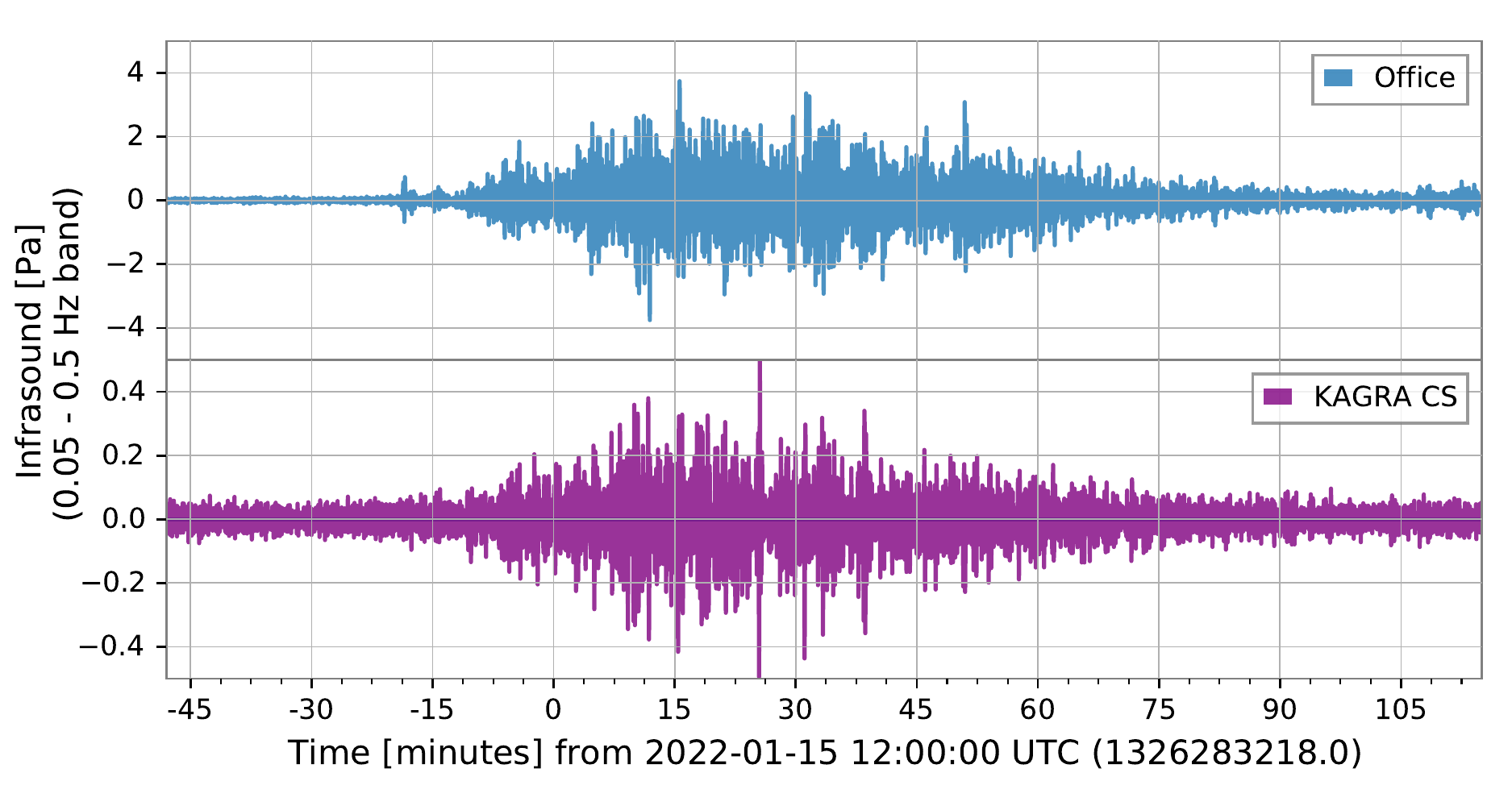}
    \caption{The time series of the infrasound sensor in the office (blue, outside of the tunnel) and in the KAGRA CS (purple, underground) with 0.05--0.5~Hz band pass filter.}
    \label{fig:infrasound}
\end{figure}

\begin{figure}[!h]\centering
    \includegraphics[width=11cm]{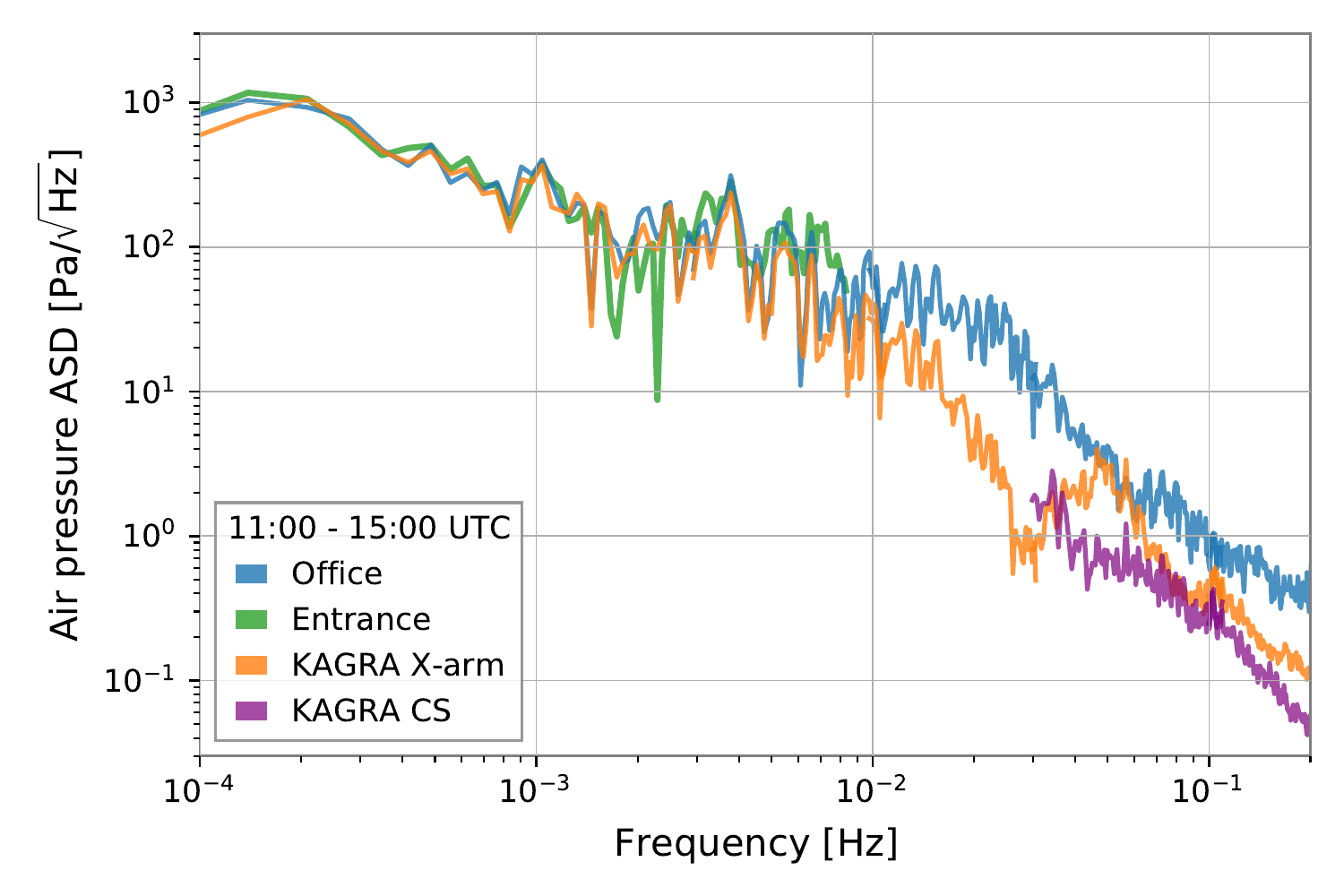}
    \caption{The ASDs of the air pressure signal (from  11:00 to 15:00 UTC) for the air pressure sensors used in this work.}
    \label{fig:ptrssureASD}
\end{figure}

The ASDs of the air pressure signal (from  11:00 to 15:00 UTC) are summarized in Fig.~\ref{fig:ptrssureASD}. In the following parts of this paper, the same time window is adopted for the analysis of air pressure signals. 
They are almost the same level below $10^{-2}$~Hz, but reduced in the tunnel over this frequency.

\subsection{Magnetic signals}
This volcanic eruption induced multiple lightning strokes, and electromagnetic waves were emitted~\cite{David, Nickolaenko}. 
The Schumann resonance, which is a global electromagnetic resonance with frequencies of 7.8~Hz, 14.1~Hz, 20.3~Hz, and so on, is generated and excited by lightning discharges in the cavity formed by the Earth's surface and ionosphere. Although the typical amplitude of the Schumann resonance is approximately 1~$\mathrm{pT/\sqrt{Hz}}$, which is less than the noise level of the used magnetometer (2~$\mathrm{pT/\sqrt{Hz}}$ for each channel), it becomes larger inside the tunnel than outside~\cite{Atsuta_2016, PhysRevD.97.102007} and is detected by the triaxial magnetometer (Bartington Mag-13MCL100~\cite{Mag13}) located at the Y-end. The same magnetometers are also located at the CS and X-end, but they are not sensitive to the Schumann resonance due to the local magnetic field noise generated by the infrastructure.

\begin{figure}[!h]\centering
    \includegraphics[width=12cm]{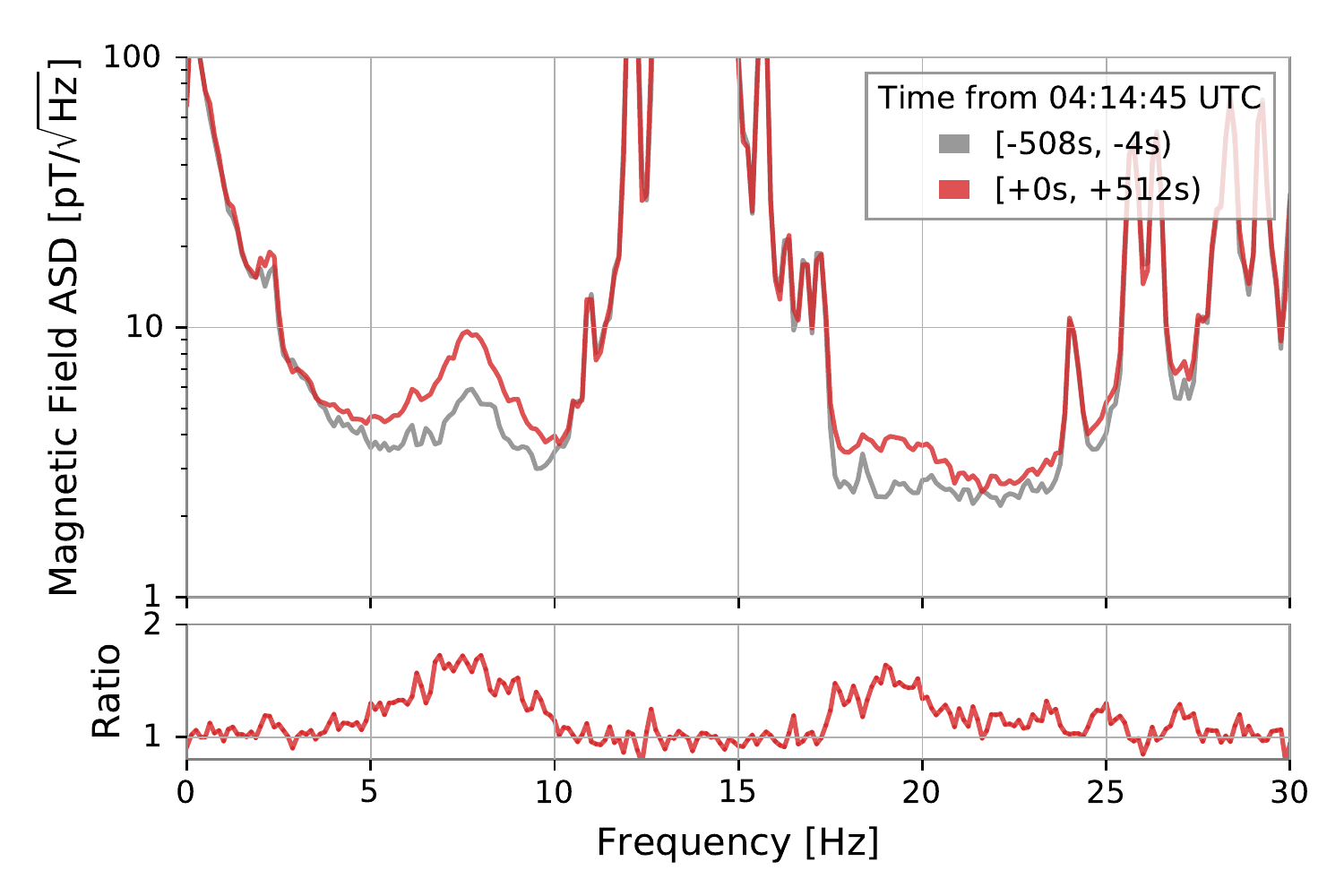}
    \caption{Top: The ASDs of the magnetometer in the KAGRA Y-end, just before (gray) and after (red) the eruption time of 04:14:45 UTC. Bottom: Their ratio.}
    \label{fig:MAG_ASD}
\end{figure}

\begin{figure}[!h]\centering
    \includegraphics[width=14cm]{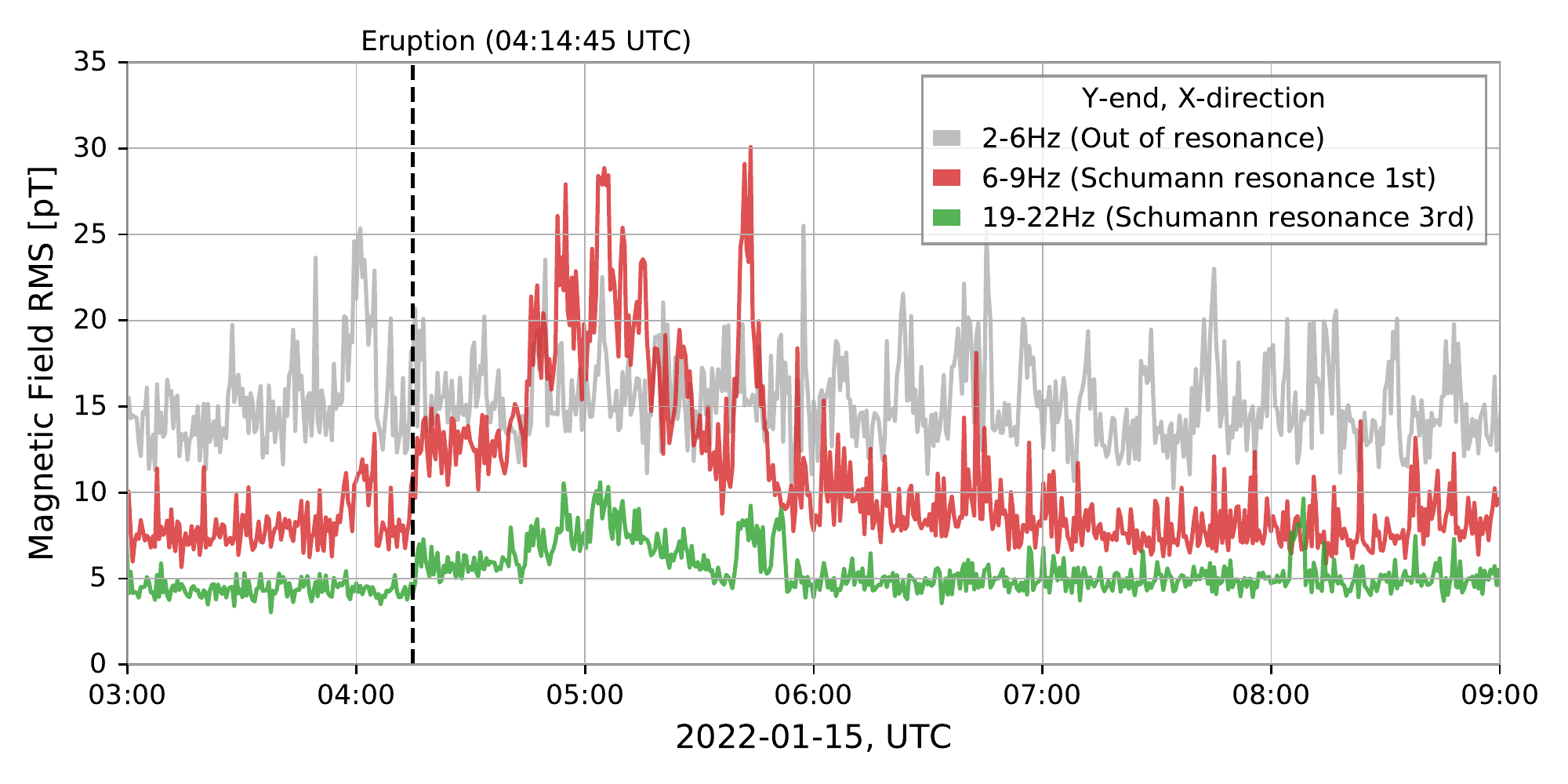}
    \caption{The time series of the RMS for the frequency bands corresponding to the 1st (red) and the 3rd (green) modes of the Schumann resonance and out of resonance (gray), for every 1~minutes. }
    \label{fig:MAG_RMS}
\end{figure}

Figure~\ref{fig:MAG_ASD} depicts the ASDs of the magnetometer at the KAGRA Y-end, just before (gray) and after (red) the eruption time of 04:14:45 UTC. The ratios are plotted in the bottom panel. 
The first mode and third mode of the Schumann resonance increased after the eruption, but other frequencies originating from the local magnetic noise did not.
Figure~\ref{fig:MAG_RMS} shows the time series of the RMS for the frequency bands corresponding to the 1st (red) and the 3rd (green) modes of the Schumann resonance and out of resonance (gray) for every 1 min. The RMS at the resonant frequencies increased immediately after the eruption without delay. This means that the magnetic signal came directly from Tonga and was not induced locally by the seismic motion (it arrived with a 10~minutes delay and its frequency was below 0.1~Hz). 
The same phenomenon was also observed at other locations, \textit{for example}, New Zealand~\cite{U09-10}.

\section{Evaluation for the transfer functions from the outside air pressure to the underground environments}
When signal $y(t)$ is completely determined by another signal $x(t)$, the linear transfer function $H(f)$ is defined by its Fourier modes, as follows:
\begin{align}
    H(f) = \frac{\tilde{y}(f)}{\tilde{x}(f)},\label{eq:TF_FFT}
\end{align}
where $\tilde{x}(f)$ and $\tilde{y}(f)$ are the (discrete) Fourier transformations of $x(t)$ and $y(t)$, respectively. 
If $y$ also contains contributions other than $x$, then the transfer function can be evaluated as:
\begin{align}
    H(f) = \frac{\braket{x(t),y(t)}}{\braket{x(t),x(t)}},\label{eq:TF_CSD}
\end{align}
where $\braket{\cdot ,\cdot }$ is the cross spectral density (CSD) of the two time series. By using the transfer function, the contribution of $x$ in $y$ can be derived as $y = H *x$ in the time domain (convolution) or in the frequency domain (multiplication).

In this section, the transfer functions from the outside air pressure to the underground environment are evaluated for the air pressure signal (from 11:00 to 15:00 UTC).

\subsection{To the underground air pressure}
\begin{figure}[!h]\centering
    \includegraphics[width=14cm]{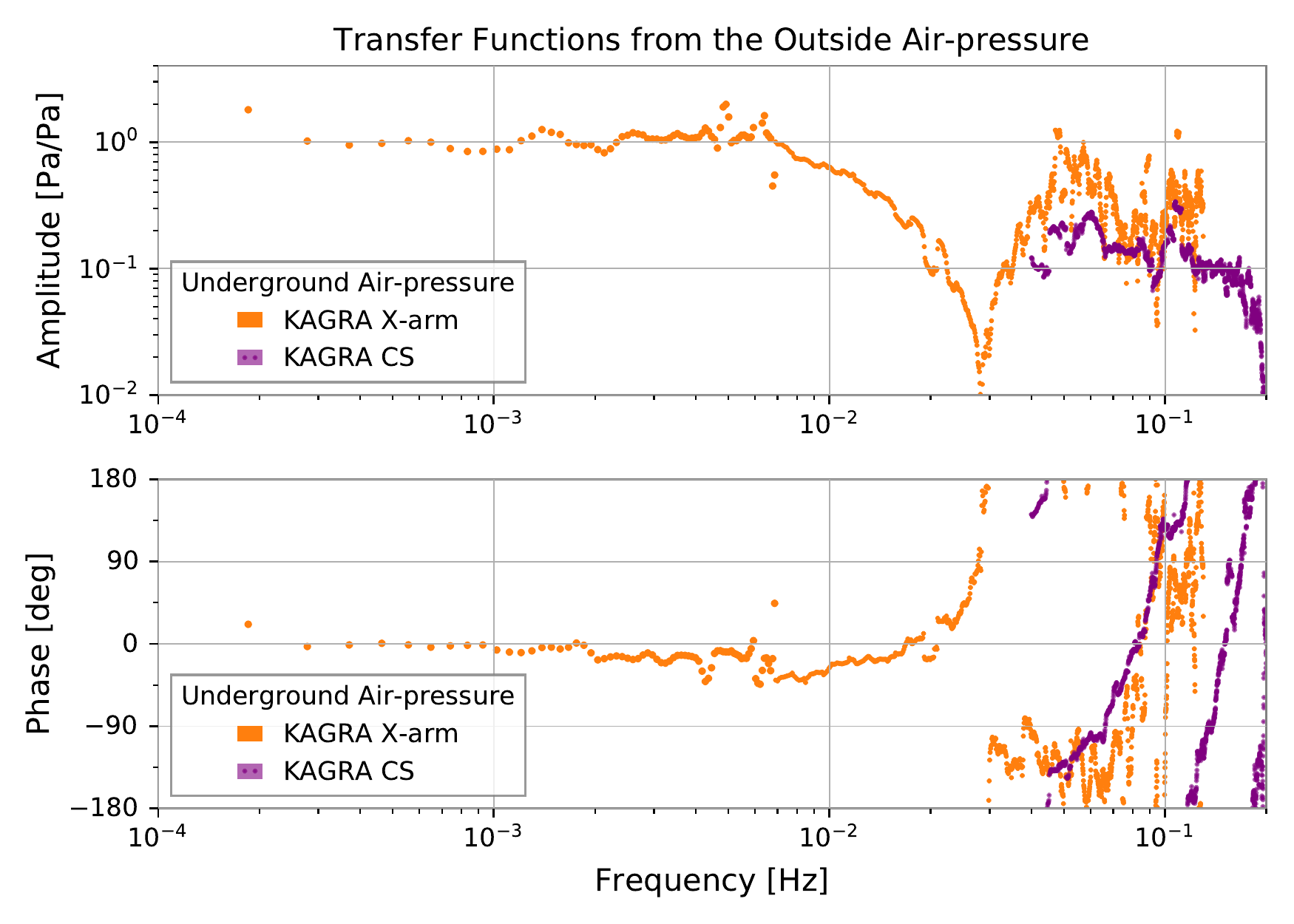}
    \caption{The bode plot of the transfer functions from the outside air pressure measured in the office to the underground air pressure measured in the X-arm (orange) and in the CS (purple).}
    \label{fig:TF_pressure}
\end{figure}

Figure~\ref{fig:TF_pressure} shows the transfer functions from the outside air pressure (in the office) to the underground air pressure (KAGRA X-arm, CS). In the X-arm, the amplitude is almost unity and the phase is almost 0\textdegree\ below 7~mHz. An anti-resonance at 28~mHz and resonant frequency at 57~mHz were observed, and these frequencies were consistent with the expected frequencies ($f=v/4L$ or $f=v/2L$) calculated from the tunnel length ($L=3$~km) and sonic speed ($v=340$~m/s). 

In the CS, the amplitude is about $1/10$ and the phase is almost proportional to the frequency with a time-shift of 14.5~seconds for 40--200~mHz. The phase plot on the linear scale and its approximation with the time-shift model are shown in Fig. ~\ref{fig:TF_infrasound_phase}. 

\subsection{To the underground seismometers}
\begin{figure}[!h]\centering
    \includegraphics[width=16cm]{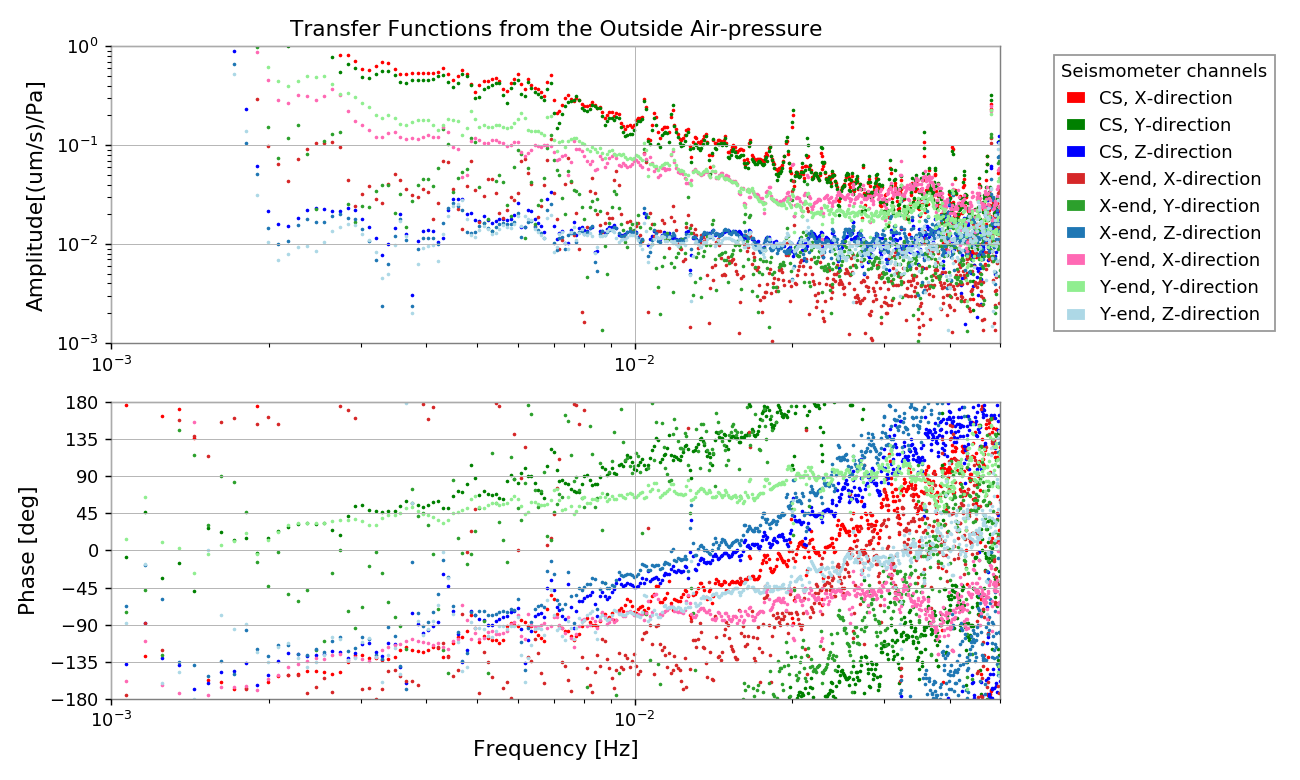}
    \caption{The bode plot of the transfer functions from the outside air pressure measured in the office to the underground seismic speeds measured in each stations of KAGRA tunnel.}
    \label{fig:TF_seis}
\end{figure}

Figure~\ref{fig:TF_seis} shows the transfer functions from the outside air pressure (in the office) to seismometers located underground. 
The red, green, and blue lines represent the X-, Y-, and Z-directions, respectively, while the deep, intermediate, and light colors represent the CS, X-end, and Y-end locations, respectively. 

The amplitude is the largest at the CS and smallest at the X-end in the horizontal  direction (X- and Y-directions) and at the same level in the vertical direction (Z-direction). 
The phase plot suggests that the motions of same direction have similar phases for all places, and the Y-direction motion is inverted with respect to the X- and Z-motions. 
The frequency dependence of the phases can be described using a time-shift model, as shown in Fig. ~\ref{fig:TF_seis_phase} for frequencies of 10--50 Hz.

\subsection{Discussion of the phase}
\begin{figure}[!h]\centering
    \includegraphics[width=10cm]{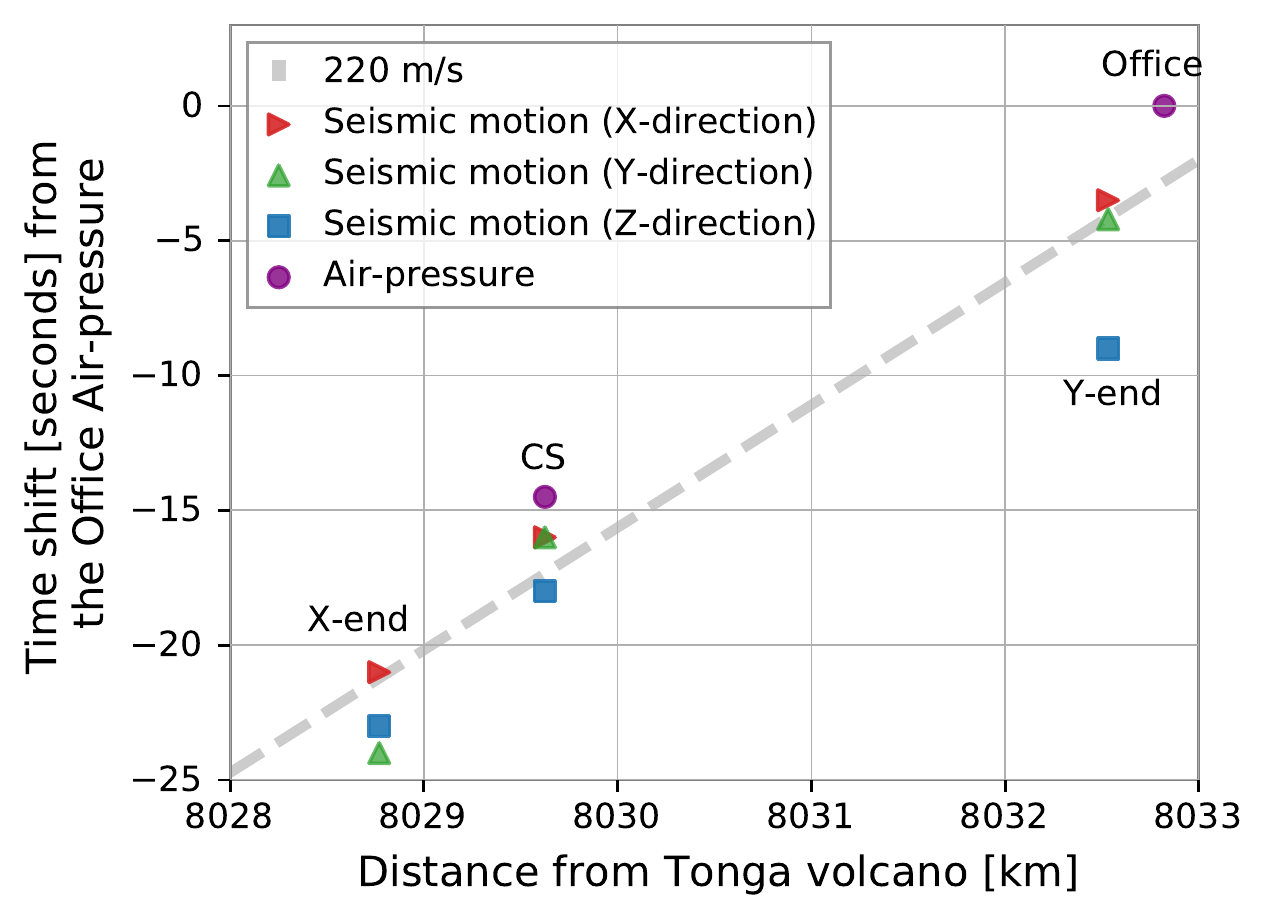}
    \caption{Relations between the distance from the Hunga Tonga-Funga Ha'apai volcano to the sensors location and the time-shift from the outside air pressure to the underground environments. The gray dashed line is corresponding to the horizontal propagation speed of 220~m/s.}
    \label{fig:TimeShift}
\end{figure}

To estimate the time shift between the outside air pressure and underground sensors from the phase of the transfer functions (Fig. ~\ref{fig:TF_infrasound_phase} and Fig. ~\ref{fig:TF_seis_phase}), they are plotted in Fig. ~\ref{fig:TimeShift} as a function of the distance from the volcano, calculated using \texttt{GRS80}. The values of the time shift were approximately proportional to the distance from the volcano, with a horizontal propagation speed of 220m/s. 
This value is similar to that for acoustic waves turning from the lower thermosphere (approximately 230 m/s) ~\cite{KUT}. 
This result suggests that the outside air pressure waves directly affect the underground environment through the mountain ground, rather than by passing through the tunnel from the entrance. 
The position dependence of the amplitude, largest at the CS and smallest at the X-end in the horizontal direction, can be understood to be caused by the topography of the mountain and the KAGRA experimental site (Figure~\ref{fig:KAGRA}), as the CS  is situated at the edge of the mountain and is easily caused by the outside air pressure. A valley exists near the Y-end. The mountain continues over the X-end, where it is more difficult to cause the ground motion.

\section{Conclusion and further prospects}\label{sec:Conclusion}
In this study, the environmental signals (seismic waves, air-pressure waves, and electromagnetic waves) caused by the volcanic eruption of the Hunga Tonga-Funga Ha'apai were monitored underground and outside of Kamioka on January 15, 2022. The transfer functions from the outside air pressure to the underground environment were evaluated, and the propagation method was discussed. 
The air pressure wave was reduced by approximately $1/10$ around 0.1~Hz but not below 0.01~Hz in the underground experimental area of KAGRA. 
It is suggested that the seismic motion is directly affected by the outside air pressure through the mountain ground rather than that passing through the tunnel from the entrance, and the level appears to be dependent on the topography. 
The position dependence of the seismic motion at the KAGRA site, which is relatively quiet at the X-end compared with the CS and Y-end, was also reported in our previous study~\cite{Hurst}. This was attributed to the X-end not having any exit tunnel, but this hypothesis turned out to be unlikely the case.

While an eruption of this magnitude is a rare event (deemed as a ''once in a century occurrence''), these results indicate the importance of monitoring the air pressure both inside and outside the tunnel, not only for transient events such as volcanic eruptions, but also normal conditions. 
These frequency regions are much lower than those of the gravitational wave signals targeted by the KAGRA. However, it is also important to consider them, because the lower-frequency vibration affects the control of suspensions and the interferometer. 
The types of infrasound sensors and barometers used in this study were different from each other, making it difficult to compare the data fairly, wherein, even cross-calibration (see Appendices A, B, and C) was performed later. For a deeper understanding of the properties of the KAGRA experimental site against air pressure waves, it is worth adding more sensors, both inside and outside the tunnel. 
A geophysical interferometer (GIF) is an asymmetric Michelson interferometer with a 1.5~km-arm and a  50~cm-arm, which has been specifically designed to measure the ground displacements (strains) along the KAGRA X-arm~\cite{PTEP3}. The analysis of the GIF data for this volcanic eruption event is also ongoing.

\medskip

This study is of benefit to future ground-based gravitational wave detectors. For example, third-generation detectors for gravitational waves with 10~km long arms, such as the Einstein Telescope (ET)~\cite{ET} in Europe and ZAIGA~\cite{ZAIGA} in China, are being planned for underground operation, as in KAGRA. 
An important inference from this study is that for the purpose of reducing seismic motions caused by atmospheric phenomena, constructing an experimental site below level ground might be better alternative to that beneath a mountain.
The other example of further plan is the torsion-bar antenna (TOBA)~\cite{TOBA1, TOBA2}, which is designed to observe gravitational waves of 10~mHz--1~Hz. The infrasound caused by a volcanic eruption overlaps with the targeted frequency range of TOBA and can directly contribute to noise. 

For these further detectors, it has been pointed out that atmospheric infrasound causes Newtonian noise and limits the sensitivity of gravitational wave observations~\cite{Donatella}. The development of infrasound monitoring techniques may be a key solution to this issue.

\section*{Acknowledgment}
This research made use of data, software, and web tools obtained or developed by the KAGRA Collaboration.
This study was supported by the KAGRA collaborators, especially administrators of the digital system and managers of the KAGRA experiment. 
We are grateful for the advice received from Prof. Masa-yuki Yamamoto (Kochi University of Technology) and Mr. Akihiro Yokota (SAYA Inc.) on infrasound sensing techniques. 
We were also provided with technical information by Mr. Kento Hiramatsu (ACO Co.Ltd.).
We would like to thank Editage (www.editage.com) for English language editing.

\medskip

The KAGRA project is funded by the Ministry of Education, Culture, Sports, Science and Technology (MEXT) and the Japan Society for the Promotion of Science (JSPS) Leading-edge Research Infrastructure Program, 
JSPS Grant-in-Aid for Specially Promoted Research 26000005, 
JSPS Grant-in-Aid for Scientific Research on Innovative Areas 2905: JP17H06358, JP17H06361 and JP17H06364, 
JSPS Core-to-Core Program A. Advanced Research Networks, 
JSPS Grant-in-Aid for Scientific Research (S) 17H06133 and 20H05639, 
JSPS Grant-in-Aid for Transformative Research Areas (A) 20A203: JP20H05854, 
the joint research program of the Institute for Cosmic Ray Research (ICRR), University of Tokyo.

This work was especially supported by the foundations, 
JSPS Grant-in-Aid for JSPS Fellows: 19J01299, JP20J21866,  
JSPS Grant-in-Aid for Early-Career Scientists: 22K14062, 
JSPS Grant-in-Aid for Research Activity Start-up: 20K22364, 
JSPS Grant-in-Aid for Scientific Research on Innovative Areas 6105: 20H05256, 22H04578, 
and the Joint Research Program of ICRR: 2019-F14, 2020-G12, 2021-G09, and 2022-G9.


\newpage
\appendix

\section{Frequency response of the sensors : Design specification}\label{sec:sensorTF}
\subsection{Trillium 120 seismometer}\label{sec:Trillium120}
\begin{figure}[!h]\centering
    \includegraphics[width=13cm]{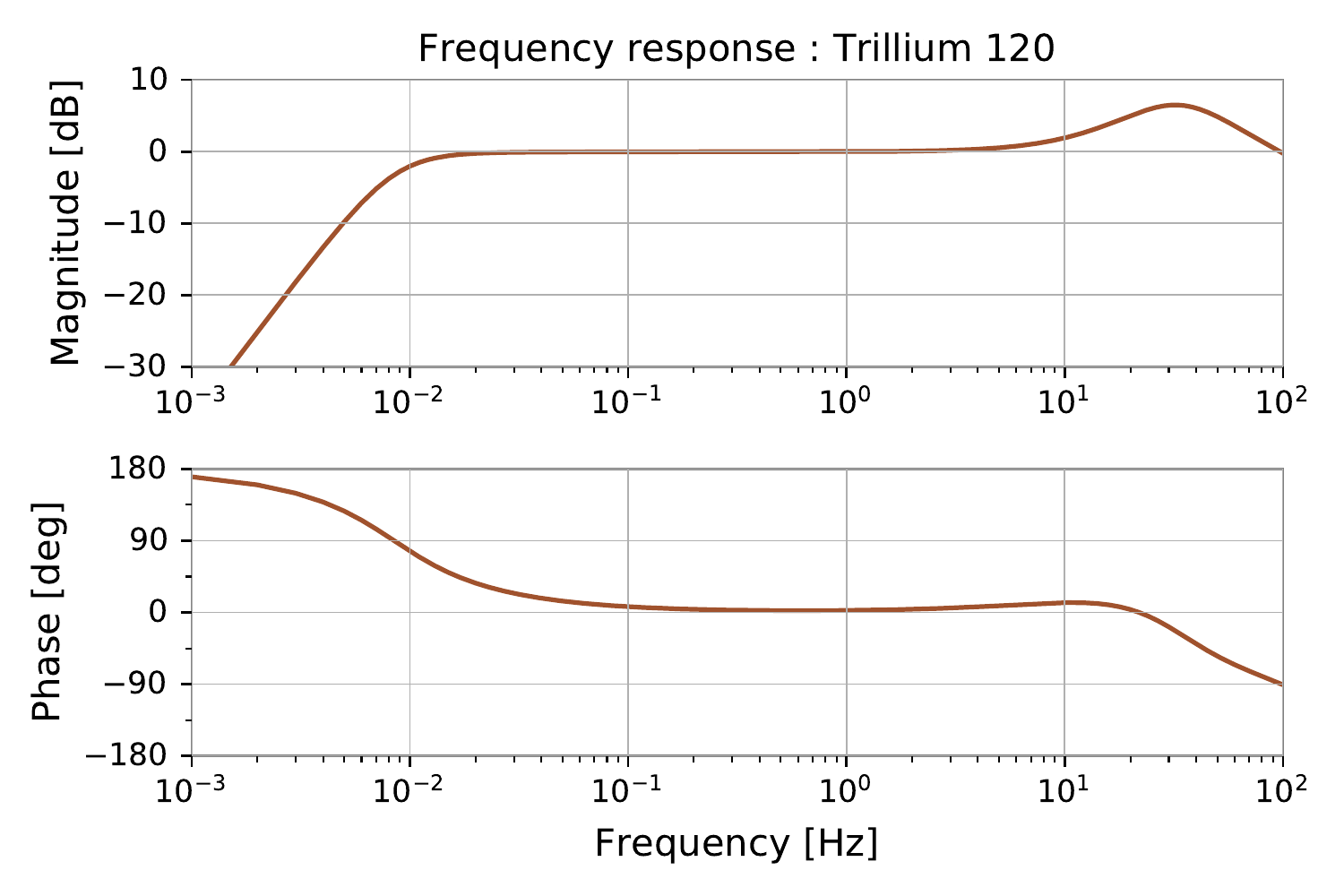}
    \caption{The bore plot of the frequency response of a Trillium120, provided by Nanometrics Inc.}
    \label{fig:Trillium120}
\end{figure}

The transfer function of an Trillium 120 from $10^{-3}$~Hz to 300~Hz is provided by Nanometrics Inc. as followings:
\begin{align}
     H(f) &=  3.080\times10^5 \times \prod_{m=1}^5 (s-z_m)  \left/ \prod_{n=1}^7 (s-p_n), \right. \\
     s &= 2\pi i f,      \nonumber\\ 
     z_1 &= z_2 = 0, \nonumber\\
     z_3 &= -90,     \nonumber\\
     z_4 &= -160.7,  \nonumber\\
     z_5 &= -3108,   \nonumber\\
     p_{1,2} &= -0.03852\pm0.03658i,  \nonumber\\
     p_3     &= -178,                 \nonumber\\
     p_{4,5} &= -135\pm160i,          \nonumber\\
     p_{6,7} &= -671\pm1154i,         \nonumber
\end{align}
Figure~\ref{fig:Trillium120} is the bode-plot. 

\newpage
\subsection{ACO 4152NHA infrasound microphone}\label{sec:TF_ACO4152NHA}
\begin{figure}[!h]\centering
    \includegraphics[width=13cm]{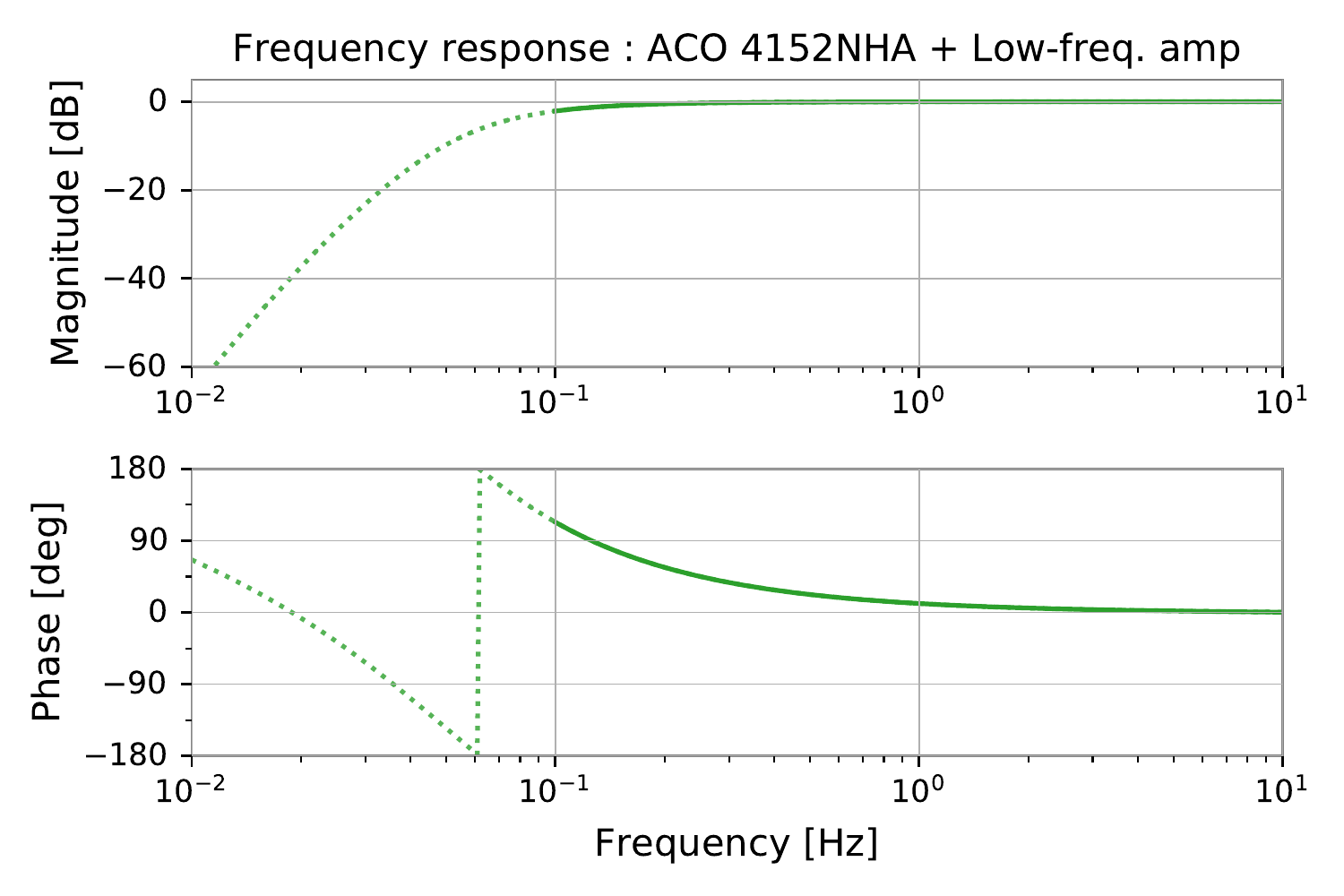}
    \caption{The bore plot of the frequency response of an ACO4152NHA, provided by ACO CO., LTD. It is extrapolated below 0.1~Hz with dotted lines. }
    \label{fig:TF_ACO4152NHA}
\end{figure}

The transfer function of an ACO 4152NHA microphone with an amplifier (ACO TYPE5006/4 low-frequency) over 0.1~Hz is provided by ACO CO., LTD. as followings:
\begin{align}
     H(f) &= 39602018.6 s^6 \left/ \sum_{n=0}^8 a_n s^n, \right. \label{eq:ACO4152NHA}\\
     s &= 2\pi i f,      \nonumber\\ 
     a_0 &= 2729.8,     \nonumber\\ 
     a_1 &= 102858.6,   \nonumber\\
     a_2 &= 1304001.5,  \nonumber\\
     a_3 &= 7990949.6,  \nonumber\\
     a_4 &= 26646639.7, \nonumber\\
     a_5 &= 48583966.6, \nonumber\\
     a_6 &= 9402975.6,  \nonumber\\
     a_7 &= 8849.2,     \nonumber\\
     a_8 &= 1.          \nonumber
\end{align}
Figure~\ref{fig:TF_ACO4152NHA} is the bode-plot. 
It is extrapolated below 0.1~Hz with dotted lines.

\newpage
\section{Cross-calibration for the microphones, infrasound sensors, and barometers}\label{sec:Calibration}
\begin{figure}[!h]\centering
    \includegraphics[width=16cm]{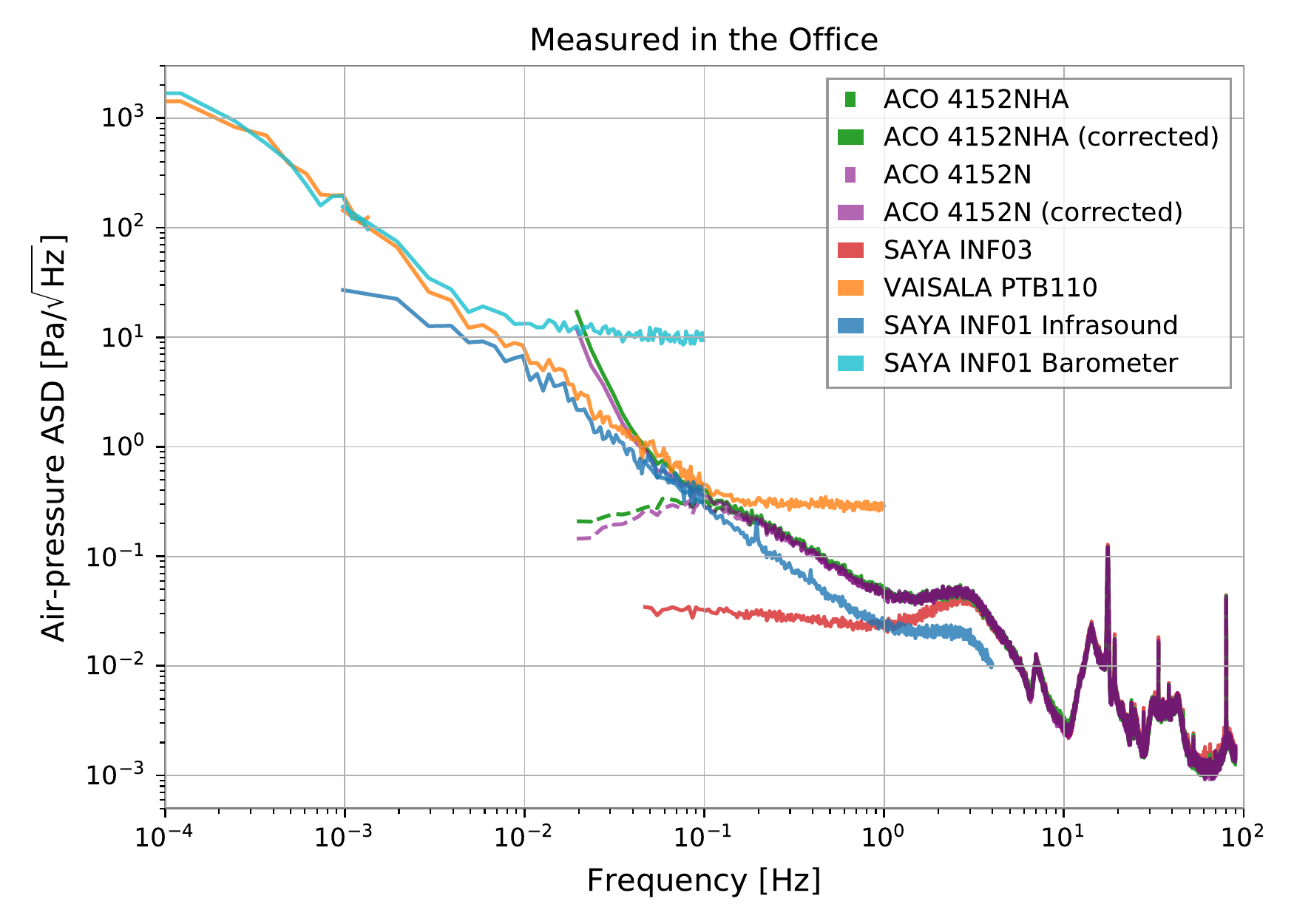}
    \caption{The ASDs for each air pressure sensor (microphones, infrasound sensors, and barometers) located close and measured simultaneously in the office. }
    \label{fig:MozumiTest_ASD}
\end{figure}

To compare the air pressure signals from the eruption observed by the different sensors, cross-calibration of all the microphones, infrasound sensors, and barometers was done in the office. Figure~\ref{fig:MozumiTest_ASD} shows the ASDs for all sensors that were located close together and measured simultaneously. 
The analog signals of ACO 4152NHA, ACO 4152N, SAYA INF03, and VAISALA PTB110 were recorded using a GRAPHTEC GL980 data logger with 200~Hz sampling. 
For ACO 4152NHA (green) and ACO 4152N (purple), the raw ASDs converted from V to Pa with constant calibration factors (1~mV/Pa and 20~mV/Pa, respectively) are plotted as dotted lines and the ASDs corrected using Eq.~(\ref{eq:ACO4152NHA}) are plotted as solid lines. 
In this section, the frequency response and noise level were evaluated using this dataset.

\subsection{ACO 4152N microphone}
\begin{figure}[!h]\centering
    \includegraphics[width=12cm]{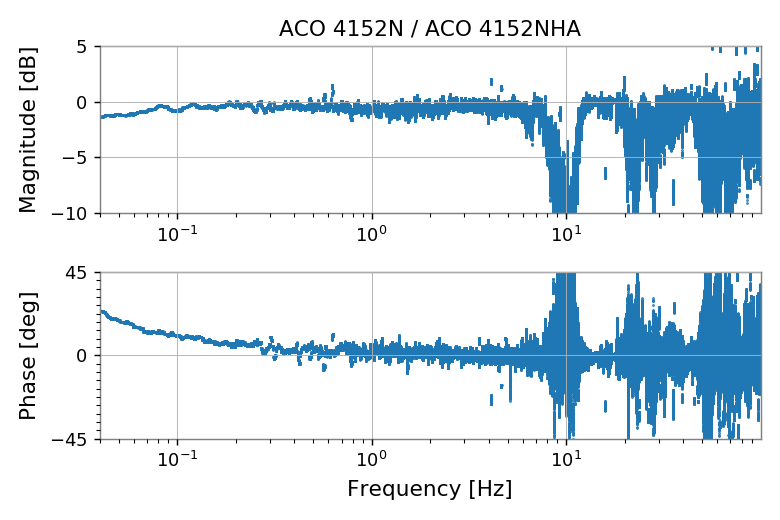}
    \caption{The bore plot of the frequency response of the ACO 4152N, evaluated based on the ACO 4152NHA.}
    \label{fig:TF_ACO4152N}
\end{figure}

The frequency response of an ACO 4152N microphone, which is used in the KAGRA CS, was evaluated based on the ACO 4152NHA infrasound microphone, and Eq.~(\ref{eq:TF_CSD}). 
Figure~\ref{fig:TF_ACO4152N} shows a Bode plot, which is almost unity, and Eq.~(\ref{eq:ACO4152NHA}) can be applied to the ACO 4152N.
The structures around 10~Hz, 20--30~Hz, and 50--70~Hz are due to the fact that the signal-to-noise ratio (SNR) of ACO 4157NHA was not good (see the next section) and was negligible.

\subsection{SAYA INF03 infrasoundsensor}
\begin{figure}[!h]\centering
    \includegraphics[width=12cm]{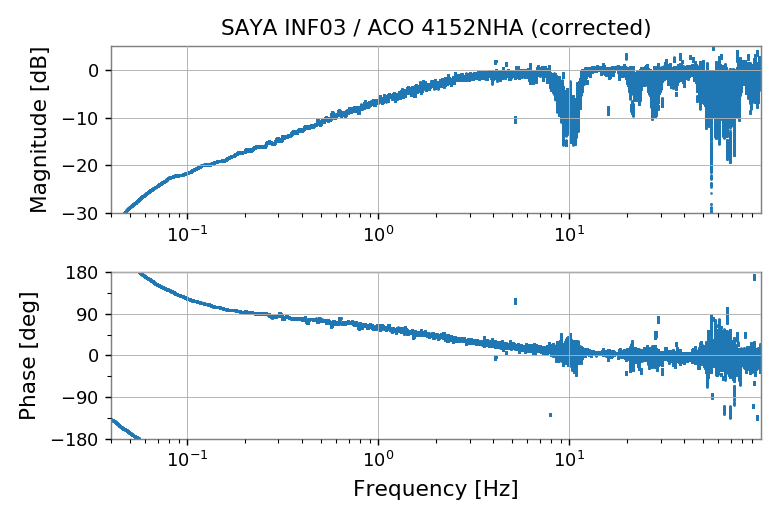}
    \caption{The bore plot of the frequency response of the SAYA INF03, evaluated based on the ACO 4152NHA.}
    \label{fig:TF_INF03}
\end{figure}

SAYA INF03 is an infrasound sensor for 0.1--1000~Hz produced by SAYA Inc. This sensor was not used for the main part of this study, but was tested for the convenience of further studies. 
The frequency response of ACO 4152NHA is shown in Fig.~\ref{fig:TF_INF03}.

\subsection{SAYA INF01 infrasoundsensor}
The frequency response of the SAYA INF01 infrasound sensor was evaluated based on the ACO 4152NHA over 0.05~Hz and based on the VAISALA PTB110 barometer below 0.1~Hz. 
As it becomes unity at $10^{-2}$--$10^{-1}$~Hz when multiplied by a factor of 1.4, this factor is applied in other sections of this paper.

\begin{figure}[!h]\centering
    \includegraphics[width=13cm]{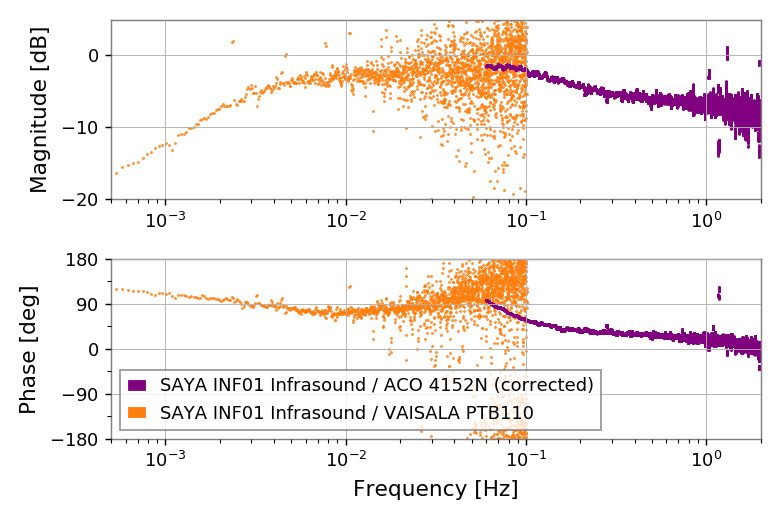}
    \caption{The bore plot of the frequency response of the SAYA INF01LE infrasound sensor, evaluated based on the ACO 4152NHA and the VAISALA PTB110.}
    \label{fig:TF_INF01}
\end{figure}

\section{Noise level of the microphones, infrasound sensors, and barometers}\label{sec:Noise}
\subsection{ACO 4152N microphone}
\begin{figure}[!h]\centering
    \includegraphics[width=10cm]{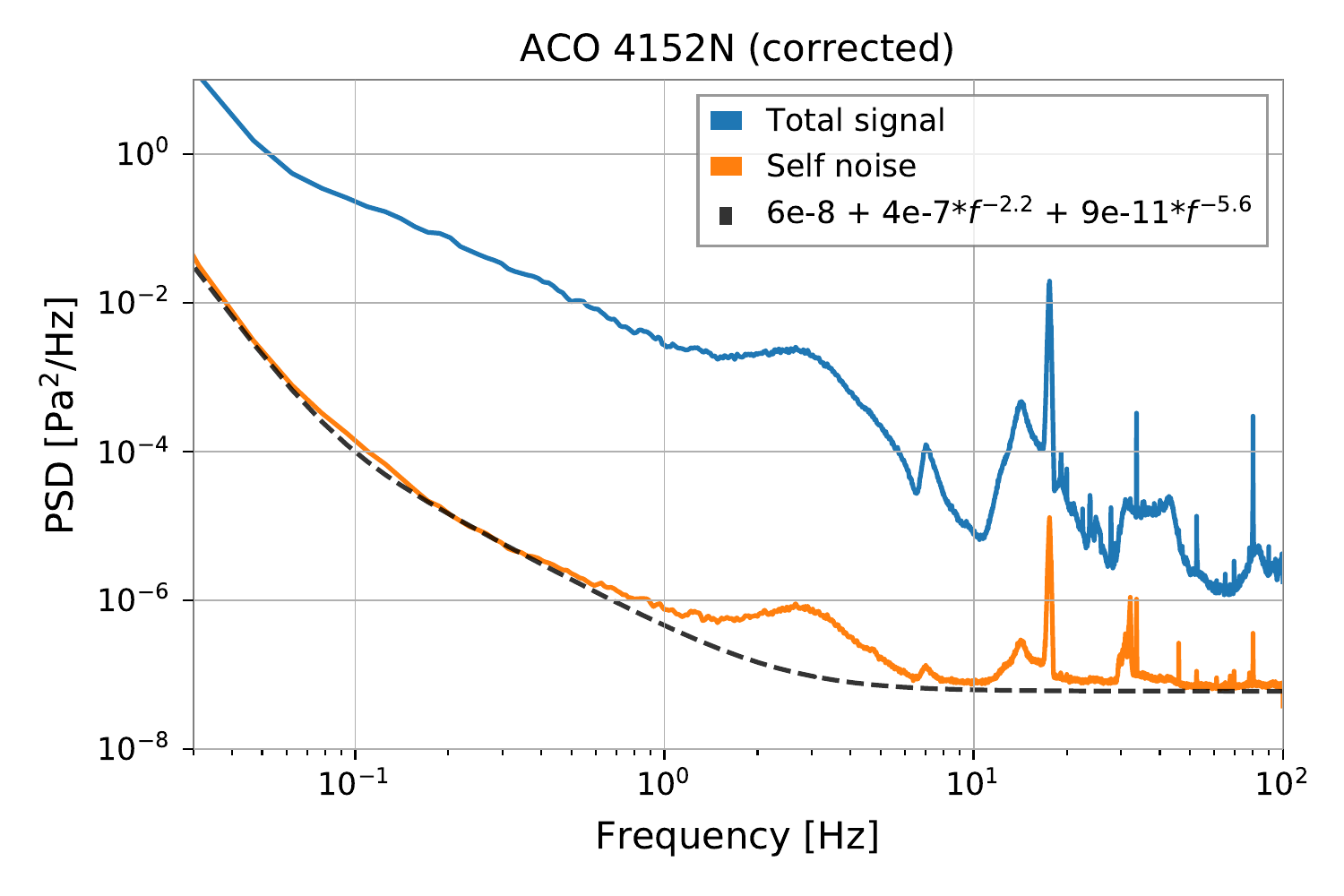}
    \caption{The PSDs of the ACO 4152N signal (blue), evaluated self noise (orange), and its approximation (black) measured in the office.}
    \label{fig:Noise_ACO4152N}
\end{figure}

The noise level of ACO 4152N is evaluated using the following procedure: 
Two sensors of the same model work in close proximity, and the differential signal of their time series is recognized as their noise ($\times 2$), with the assumption that their individual differences are small. 
The orange line in Fig.~\ref{fig:Noise_ACO4152N} shows the result, and the black dotted line represents its approximation.

\subsection{ACO 4152NHA and SAYA INF03 infrasound microphones}
The PSD of the noise level $P_{y, \mathrm{noise}}(f)$ in a sensor $y(t)$ can be evaluated using two reference sensors $x_1(t), x_2(t)$ when their SNR are sufficiently good, 

\begin{align}
    P_{y, \mathrm{noise}}(f) &= P_{y}(f) -  \left|\frac{\braket{x_1(t),y(t)}\ \braket{y(t),x_2(t)}}{\braket{x_1(t),x_2(t)}} \right|, \\
    P_{y}(f) &= \braket{y(t),y(t)}.
\end{align}

\begin{figure}[!h]\centering
    \includegraphics[width=10cm]{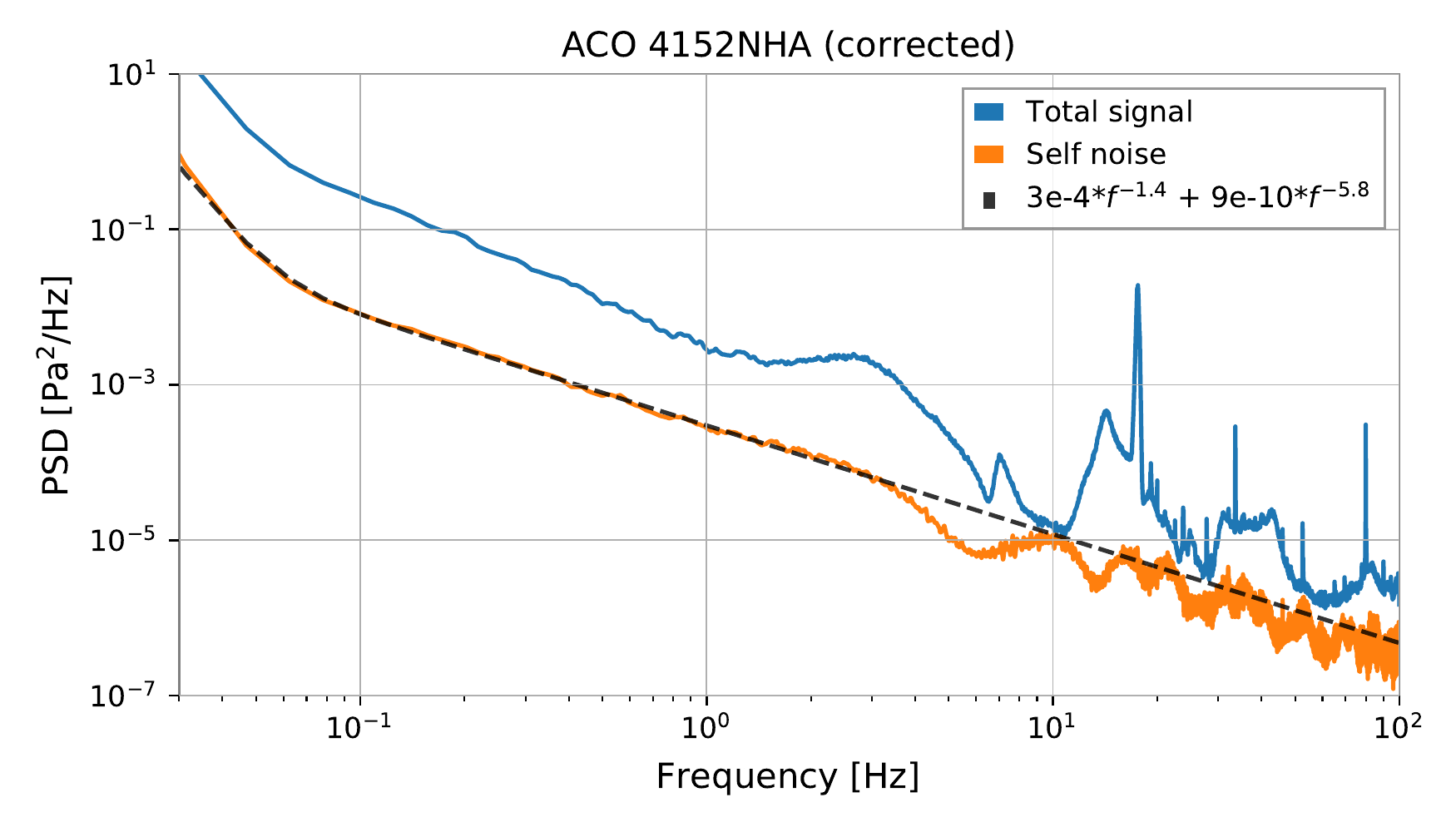}\\
    \includegraphics[width=10cm]{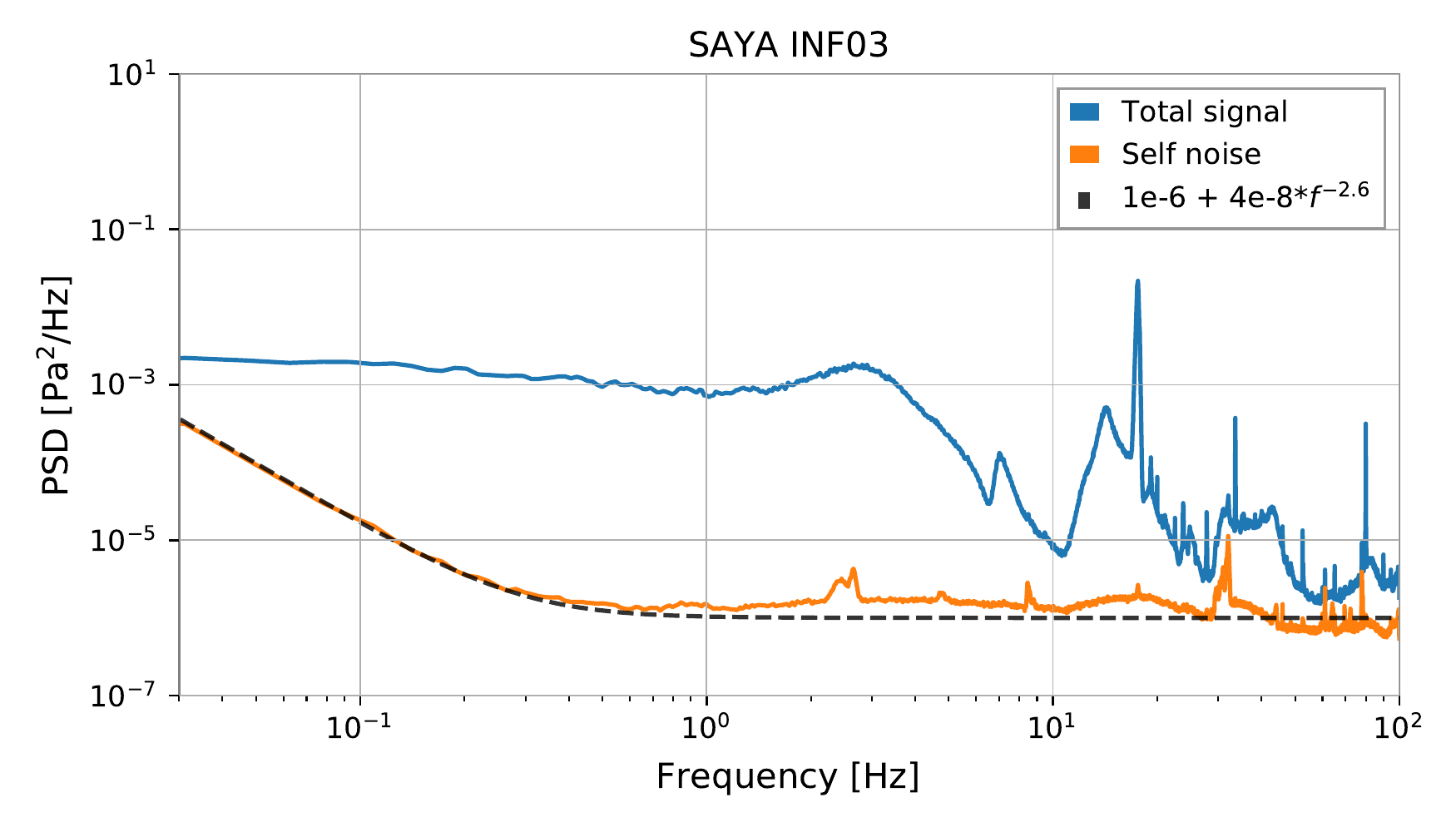}
    \caption{The PSDs of the ACO 4152NHA (top) and the SAYA INF03 (bottom) signals (blue), evaluated self noises (orange), and their approximation (black) measured in the office.}
    \label{fig:Noise_INF03}
\end{figure}

Figure~\ref{fig:Noise_INF03} shows the self-noise results of ACO 4152NHA and SAYA INF03 evaluation using this method and two references (ACO 4152N).

\subsection{SAYA INF01LE infrasound sensor}
\begin{figure}[!h]\centering
    \includegraphics[width=12cm]{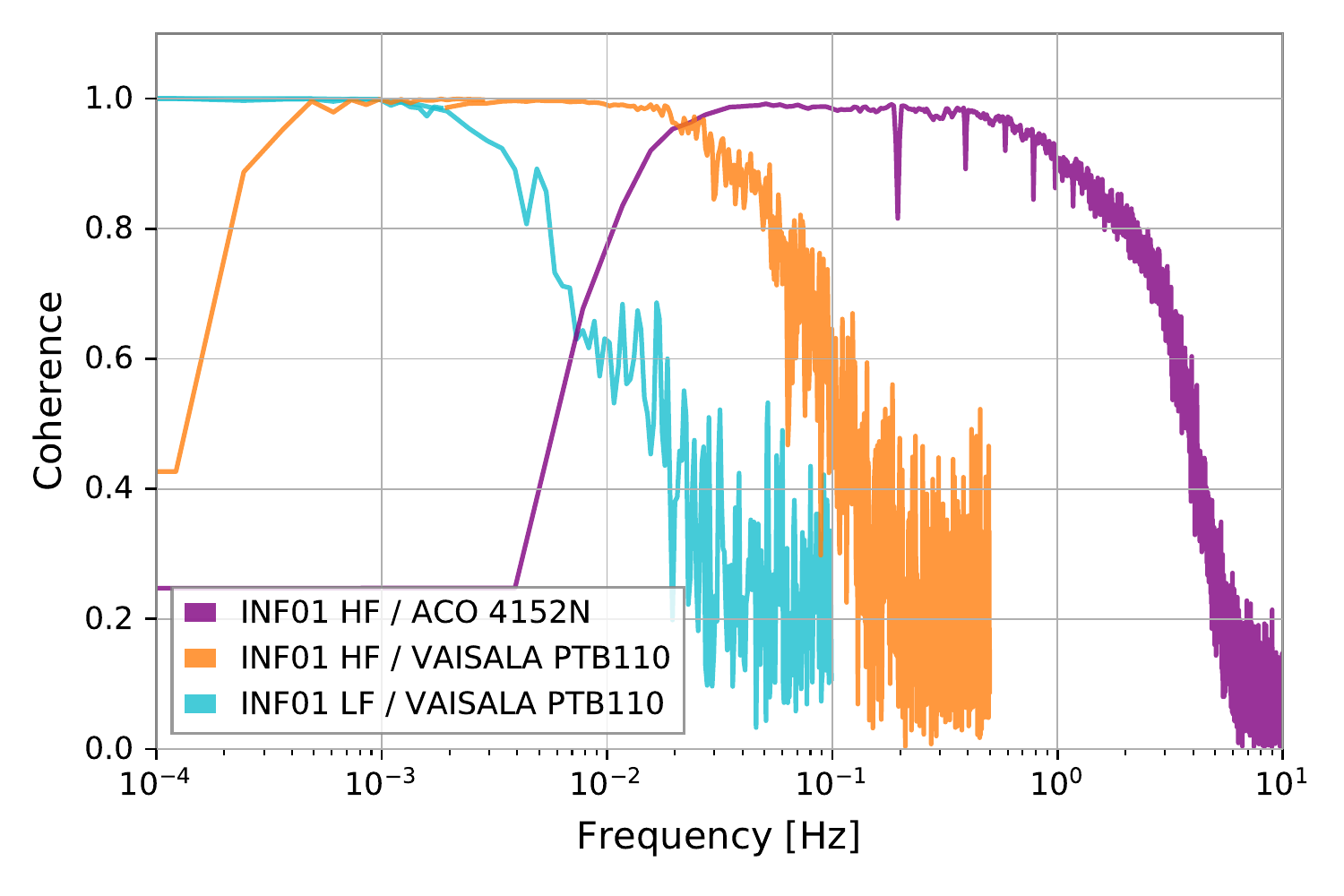}\\
    \includegraphics[width=12cm]{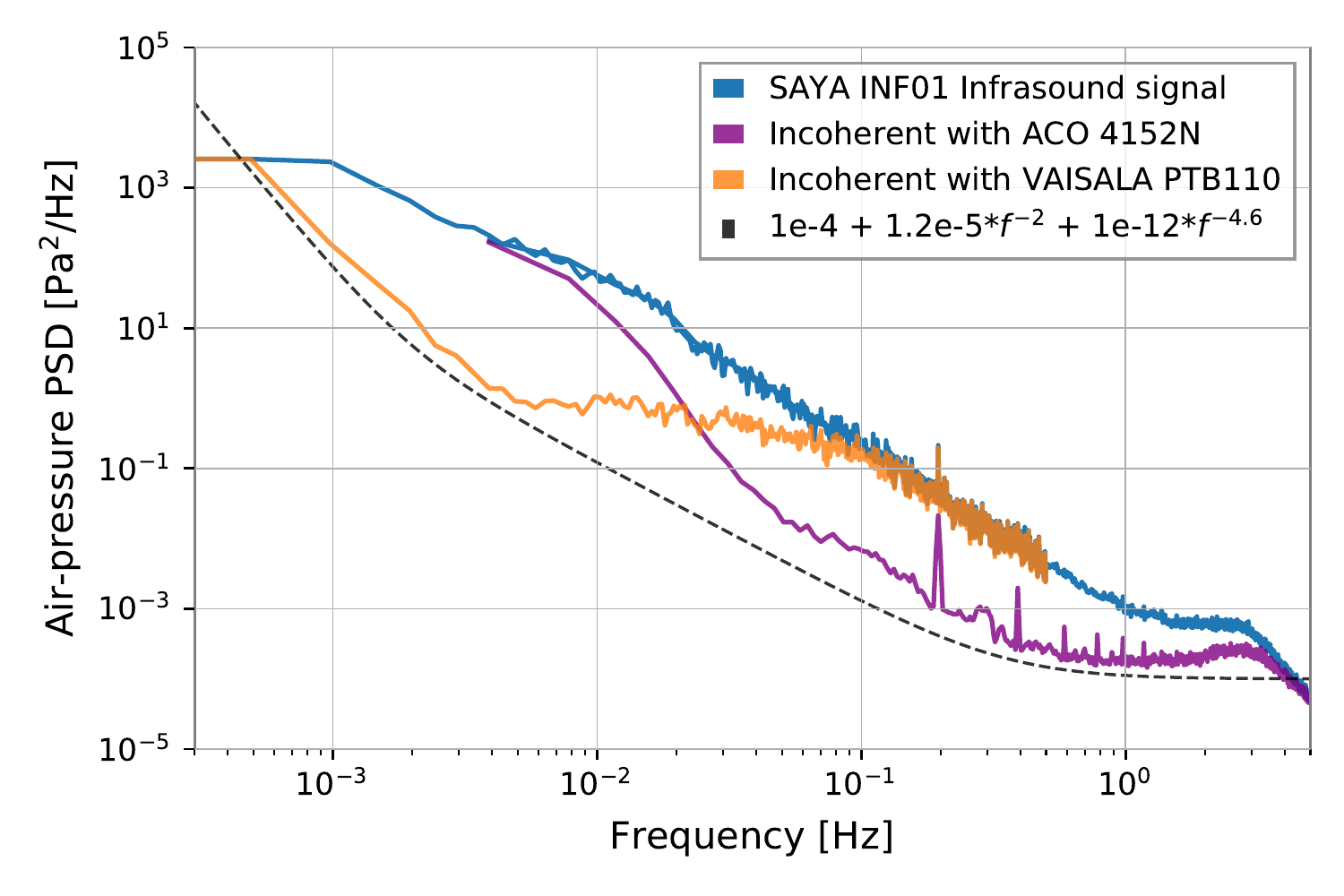}
    \caption{Top: Coherence between the SAYA INF01LE and ACO 4152N or VAISALA PTB110. 
    Bottom: The PSDs of the SAYA INF01LE signal (blue), evaluated incoherent signal with ACO 4152N (purple) and with VAISALA PTB110 (orange), and the approximated self noise (black) measured in the office.}
    \label{fig:Noise_INF01}
\end{figure}

Because there are no good reference sensors covering the frequency range of the SAYA INF01 infrasound sensor, the coherence (defined in Eq.(\ref{eq:coherence})) was used to estimate the self-noise. 
In the PSD of the SAYA INF01LE infrasound sensor signal $P(f)$, the components that are coherent with another sensor can be written as $\gamma^2(f)\cdot P(f)$ and the remaining components $[1-\gamma^2(f)]\cdot P(f)$ are incoherent. The envelope of the incoherent components can be understood as self-noise. 
Figure~\ref{fig:Noise_INF01} shows the result of the self-noise estimation for the SAYA INF01LE infrasound sensor.

\newpage
\section{Coherence}
The coherence $\gamma(f)$ for two time series $x(t), y(t)$ is defined as
\begin{align}
    \gamma_{x,y}^2(f) = \frac{|\braket{x(t),y(t)}|^2}{\braket{x(t),x(t)}\braket{y(t),y(t)}}, \label{eq:coherence}
\end{align}
and it shows the ratio of the common signal in $x$ and $y$.

\begin{figure}[!h]\centering
    \includegraphics[width=12cm]{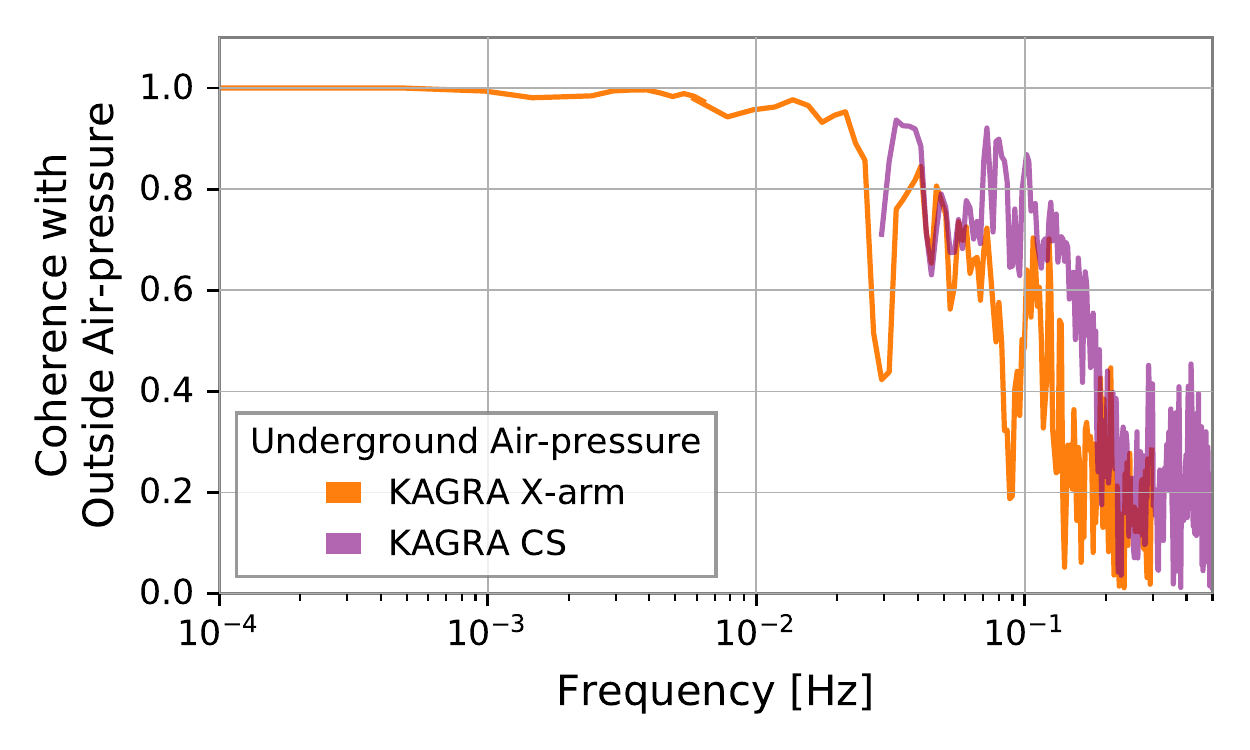}
    \caption{The coherence of the air pressure between inside and outside of the tunnel for the eruption signal.}
    \label{fig:coherence_INF-pressure}
\end{figure}

\begin{figure}[!h]\centering
    \includegraphics[width=14cm]{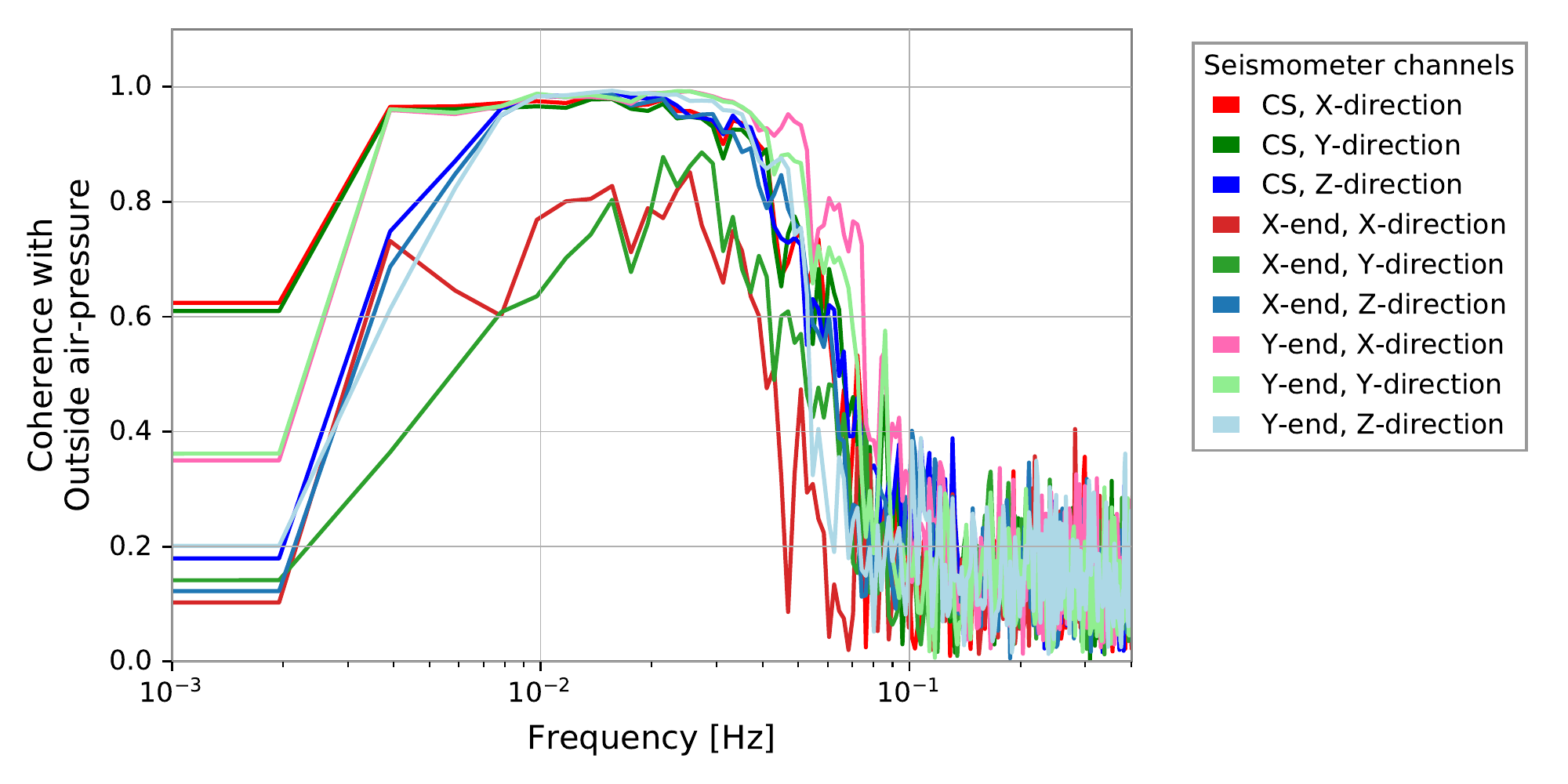}
    \caption{The coherence between the outside air pressure and the underground seismic motions for the eruption signal.}
    \label{fig:coherence_INF-seis}
\end{figure}

\newpage
\section{TF phase and Time shift model}

\begin{figure}[!h]\centering
    \includegraphics[width=11cm]{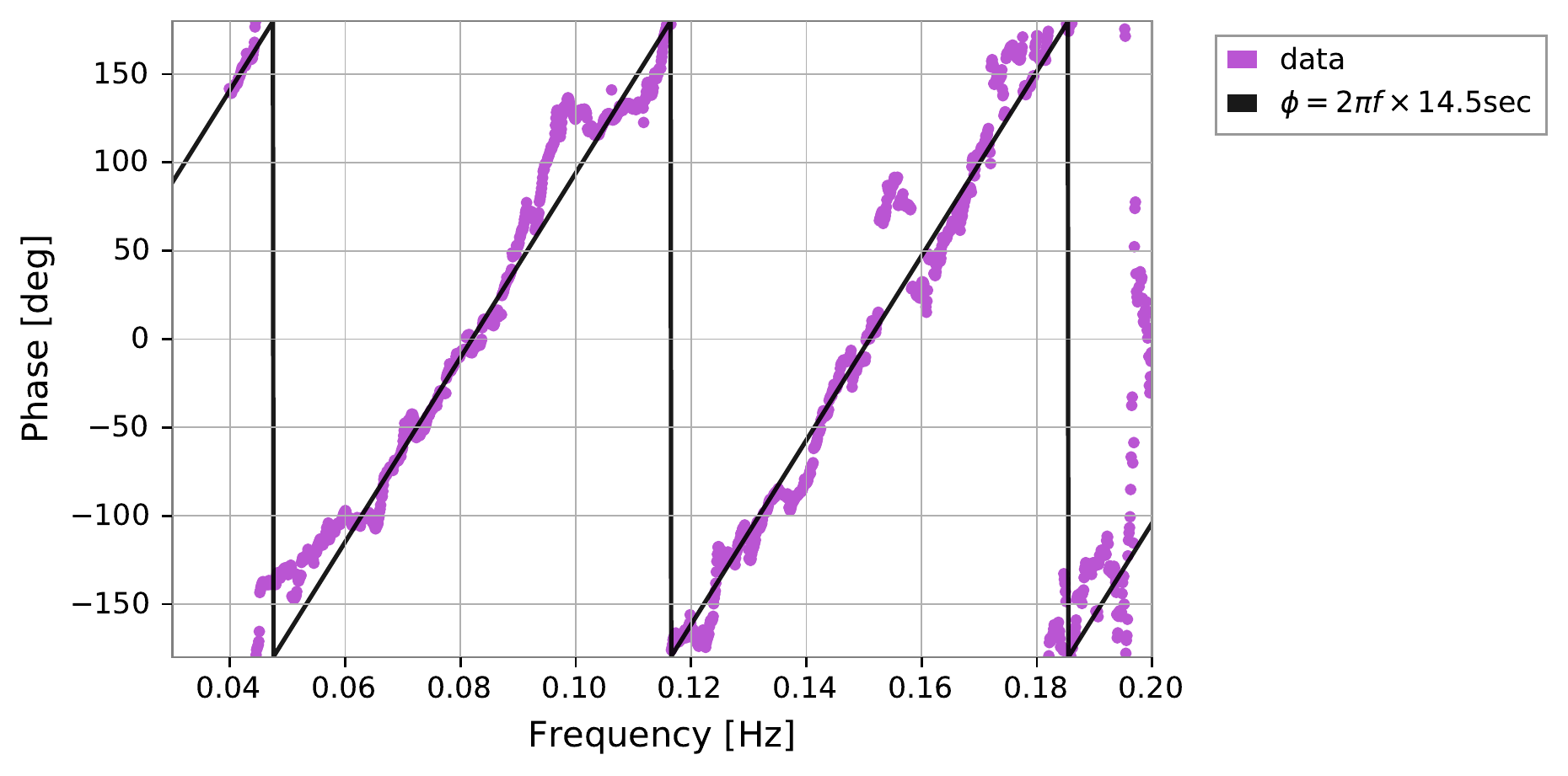}
    \caption{The phase of the transfer function of air pressure from the outside to KAGRA CS in the liner scale.}
    \label{fig:TF_infrasound_phase}
\end{figure}

\begin{figure}[!h]\centering
    \includegraphics[width=16cm]{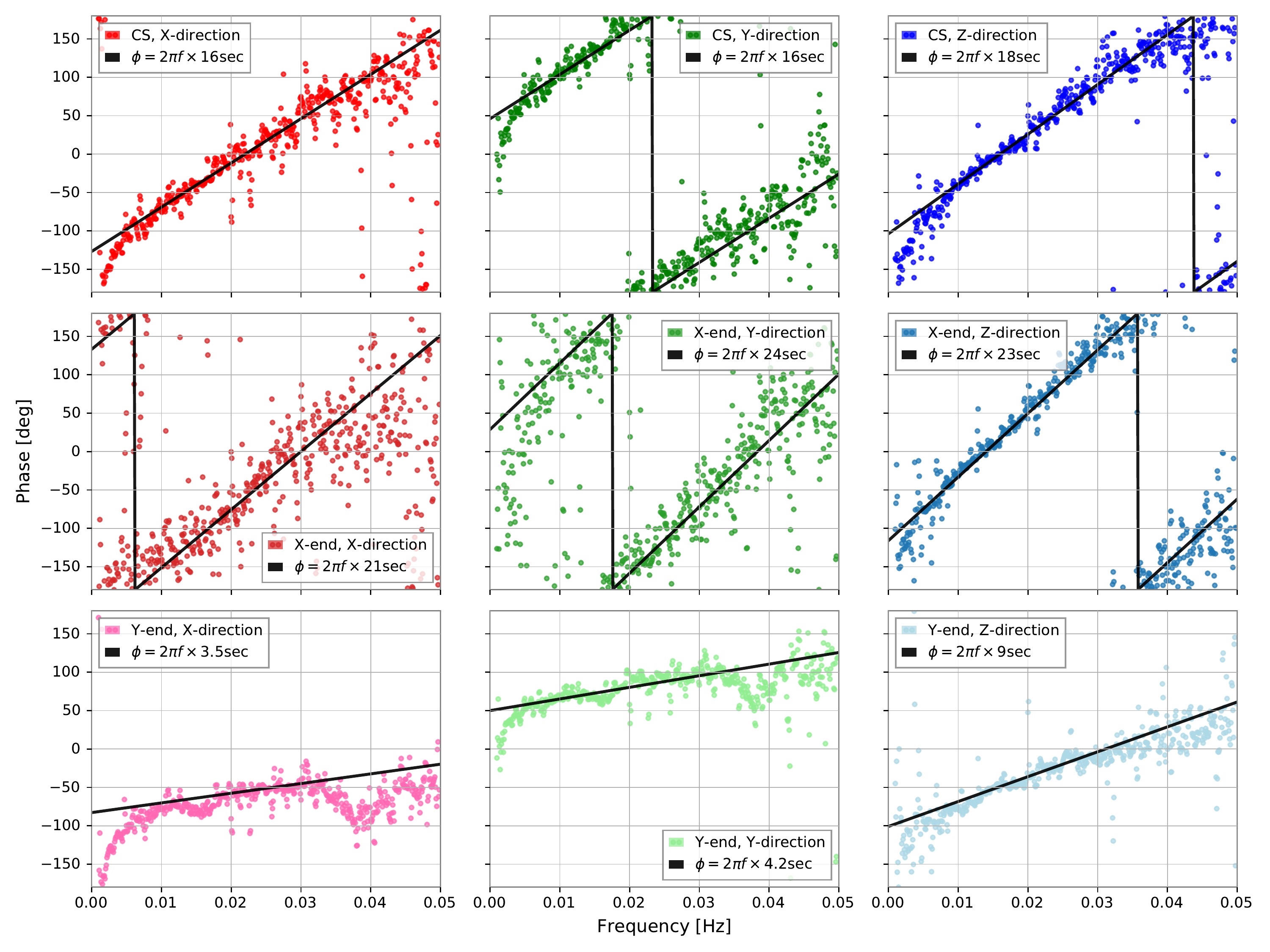}
    \caption{The phase of the transfer function from the outside air pressure to the underground seismic motions in the liner scale.}
    \label{fig:TF_seis_phase}
\end{figure}


\begin{thebibliography}{99}


\bibitem{PTEP01}
T. Akutsu et al. (KAGRA collaboration),
"\emph{Overview of KAGRA : Detector design and construction history}",
Prog. Theor. Exp. Phys., Volume 2021, Issue 5, 05A101. 
[\url{https://doi.org/10.1093/ptep/ptaa125}]

\bibitem{Galaxies}
H. Abe et al. (KAGRA collaboration),
"\emph{The Current Status and Future Prospects of KAGRA, the Large-Scale Cryogenic Gravitational Wave Telescope Built in the Kamioka Underground}",
Galaxies 2022, 10, 63. 
[\url{https://doi.org/10.3390/galaxies10030063}]


\bibitem{LIGO}
The LIGO Scientific Collaboration,
"\emph{Advanced LIGO}",
Class. Quantum Gravity 2015, 32, 074001. 
[\url{https://doi.org/10.1088/0264-9381/32/7/074001}]

\bibitem{Virgo}
F. Acernese et al., 
"\emph{Advanced Virgo: A second-generation interferometric gravitational wave detector}",
Class. Quantum Gravity 2014, 32, 024001. 
[\url{https://doi.org/10.1088/0264-9381/32/7/074001}]

\bibitem{GEO}
K. Dooley et al., 
"\emph{GEO 600 and the GEO-HF upgrade program: Successes and challenges}",
Class. Quantum Gravity 2015, 33. 
[\url{https://doi.org/10.1088/0264-9381/33/7/075009}]

\bibitem{Tomaru}
T. Tomaru et al.,
“\emph{Vibration-Free Pulse Tube Cryocooler System for Gravitational Wave Detectors, Part I: Vibration-Reduction Method and Measurement}",
Cryocoolers 13. pp 695-702 (2005). 
[\url{https://doi.org/10.1007/0-387-27533-9_86}]

\bibitem{PTEP3} 
T. Akutsu et al. (KAGRA Collaboration), 
"\emph{Overview of KAGRA: Calibration, detector characterization, physical environmental monitors, and the geophysics interferometer}", 
Prog. Theor. Exp. Phys., Volume 2021, Issue 5, 05A102.
[\url{https://doi.org/10.1093/ptep/ptab018}]

\bibitem{USGS} 
"\emph{M 5.8 Volcanic Eruption - 68 km NNW of Nuku‘alofa, Tonga}", 
The USGS Earthquake Hazards Program of the U.S. Geological Survey (USGS).
[\url{https://earthquake.usgs.gov/earthquakes/eventpage/us7000gc8r/origin/detail}]

\bibitem{GVP} 
Global Volcanism Program, (Crafford, A.E., and Venzke, E., eds.), 
"\emph{Report on Hunga Tonga-Hunga Ha'apai (Tonga)}", 
S.K. Sennert (Ed.), Weekly Volcanic Activity Report, 19 January-25 January 2022, Smithsonian Institution and US Geological Survey (2022)
[\url{https://volcano.si.edu/showreport.cfm?doi=10.5479/si.GVP.BGVN202202-243040}]

\bibitem{Klein} 
A. Klein,
"\emph{Tongan volcano erupts}",
New Scientist, Volume 253, Issue 3370, 2022, Page 7.
[\url{https://doi.org/10.1016/S0262-4079(22)00074-4}]

\bibitem{Zhao} 
W. Zhao, C. Sun, Zh. Guo, 
"\emph{Reawaking of Tonga volcano}", 
The Innovation, Volume 3, Issue 2, 2022, 100218. 
[\url{https://doi.org/10.1016/j.xinn.2022.100218}]

\bibitem{KUT} 
Y. Nishikawa, M.-Y. Yamamoto, K. Nakajima, I. Hamama, H. Saito, Y. Kakinami, 
"\emph{What excited tsunami from Tonga 2022 eruption? Observation and theory}",
[\url{https://doi.org/10.21203/rs.3.rs-1513574/v1}]

\bibitem{Kataoka} 
R. Kataoka , S. D. Winn, E. Touber, 
"\emph{Meteotsunamis in Japan associated with the Tonga Eruption in January 2022}",
[\url{https://doi.org/10.31223/X55K8V}]

\bibitem{Saito} 
S. Saito,  
"\emph{Ionospheric disturbances observed over Japan following the eruption of Hunga Tonga-Hunga Ha’apai on 15 January 2022. Earth Planets Space 74, 57 (2022)}", 
Earth, Planets and Space volume 74, Article number: 57 (2022) 
[\url{https://doi.org/10.1186/s40623-022-01619-0}]


\bibitem{T120QA} 
Nanometrics Inc., 
"\emph{Trillium 120 Q/QA}",
[\url{https://www.nanometrics.ca/our-science/trillium-120-qqa-0}]

\bibitem{ObsPy} 
M. Beyreuther, R. Barsch, L.Krischer, T. Megies, Y. Behr, J. Wassermann
"\emph{ObsPy: A Python Toolbox for Seismology}",
Seismological Research Letters (2010) 81 (3): 530–533.
[\url{ttps://doi.org/10.1785/gssrl.81.3.530}]

\bibitem{U09-P29} 
K. Tarumi, K. Yoshizawa, 
"\emph{Global propagation of seismic waves generated by the explosive eruptions of Hunga Tonga-Hunga Ha’apoi: Preliminary analyses}, 
Japan Geoscience Union Meeting 2022, U09-P29

\bibitem{Peterson}
J. R. Peterson,
"\emph{Observations and modeling of seismic background noise}, 
U.S. Geol. Surv. Tech. Rept. (1993), 93-322, 1–95. 
[\url{https://doi.org/10.3133/ofr93322}]

\bibitem{SAYA} 
M.-Y. Yamamoto, A. Yokota,  
"\emph{Infrasound monitoring for disaster prevention from geophysical destructions}", 
in Proceedings of the 5th International Symposium on Frontier Technology 24–28 (Kunming, China, 2015).

\bibitem{SAYA-INF01LE} 
SAYA Inc.,
"\emph{Infrasound Sensor ADX3I-INF01LE. Multifunction-I/O XIII}", 
[\url{https://www.sayanet.com/products/INF01.html}]

\bibitem{CAGmon}
P. Jung et al.,
"\emph{Identifying and diagnosing coherent associations and causalities between multi-channels of the gravitational wave detector}", 
Phys. Rev. D 106, 042010 (2022). 
[\url{https://link.aps.org/doi/10.1103/PhysRevD.106.042010}]

\bibitem{David} 
A. David et al., 
"\emph{Under the surface: Pressure-induced planetary-scale waves, volcanic lightning, and gaseous clouds caused by the submarine eruption of Hunga Tonga-Hunga Ha'apai volcano}", 
Earthquake Research Advances, 2022, 100134,
[\url{https://doi.org/10.1016/j.eqrea.2022.100134}]

\bibitem{Nickolaenko} 
A. Nickolaenko, A. Y. Schekotov, M. Hayakawa, R. Romero, J. Izutsu,  
"\emph{Electromagnetic Manifestations of Tonga Eruption in Schumann Resonance Band}", 
Available at SSRN. 
[\url{http://dx.doi.org/10.2139/ssrn.4051361}]

\bibitem{Atsuta_2016}
S. Atsuta et al., 
"\emph{Measurement of Schumann Resonance at Kamioka}", 
J. Phys. 2016, 716, 012020.
[\url{https://doi.org/10.1088/1742-6596/716/1/012020}]

\bibitem{PhysRevD.97.102007}
M.W. Coughlin et~al., 
"\emph{Measurement and subtraction of Schumann resonances at gravitational-wave interferometers}",
Phys. Rev. D 2018, 97, 102007.
[\url{https://doi.org/10.1103/PhysRevD.97.102007}]

\bibitem{Mag13}
Bartington Instruments Ltd, Mag-13 Three-axis, 
[\url{https://www.bartington.com/products/precision-magnetometers/mag-13-three-axis/}]

\bibitem{U09-10}
T. G. Caldwell, P. Jarvis, C. Noble and Y. Ogawa’  rather than  ‘C. Grant, N. Chris, J. Paul, O. Yasuo, 
"\emph{Constraints on the Hunga Tonga-Hunga Ha'apai eruption timing from remote magnetic field measurements}", 
Japan Geoscience Union Meeting 2022, U09-10


\bibitem{Hurst}
A. Longo, S. Bianchi, W. Plastino, K. Miyo, T. Yokozawa, T. Washimi and A. Araya,
"\emph{Local Hurst Exponent Computation of Data from Triaxial Seismometers Monitoring KAGRA}”,  
Pure Appl. Geophys. 178, 3461–3470 (2021). 
[\url{https://doi.org/10.1007/s00024-021-02810-2}]

\bibitem{ET}
M. Punturo et al., 
“\emph{The Einstein Telescope: A third-generation gravitational wave observatory}”,  
Class. Quant. Grav. 27 (2010) 194002.
[\url{https://doi.org/10.1088/0264-9381/27/19/194002}]

\bibitem{ZAIGA}
M. S. Zhan et al., 
"\emph{ZAIGA: Zhaoshan Long-baseline Atom Interferometer Gravitation Antenna}", 
International Journal of Modern Physics, Vol. 29, No.~4 (2020) 1940005.
[\url{https://doi.org/10.1142/S0218271819400054}]

\bibitem{TOBA1} 
M. Ando et al., 
"\emph{Torsion-Bar Antenna for Low-Frequency Gravitational-Wave Observations}", 
Phys. Rev. Lett., 105, 16, 161101, 4, (2010). 
[\url{https://link.aps.org/doi/10.1103/PhysRevLett.105.161101}]

\bibitem{TOBA2} 
T. Shimoda, S. Takano, C. P. Ooi, N. Aritomi, Y. Michimura, M. Ando, 
"\emph{Torsion-Bar Antenna: A ground-based mid-frequency and low-frequency gravitational wave detector}", 
International Journal of Modern Physics D, Vol. 29, No. 4 (2020) 1940003.
[\url{https://doi.org/10.1142/S0218271819400030}]

\bibitem{Donatella}
D. Fiorucci, J. Harms, M. Barsuglia, I. Fiori, and F. Paoletti, 
"\emph{Impact of infrasound atmospheric noise on gravity detectors used for astrophysical and geophysical applications}
Phys. Rev. D 97, 062003, (2018)
[\url{https://doi.org/10.1103/PhysRevD.97.062003}]

\end{thebibliography}
\end{document}